\documentclass[11pt]{article}
\usepackage{indentfirst}
\usepackage{comment}
\usepackage{appendix}
\usepackage{graphicx}
\usepackage{amsmath}
\numberwithin{equation}{section}
\usepackage{amsfonts}
\usepackage{amssymb}
\usepackage{setspace}
\usepackage{booktabs}
\usepackage{cancel}
\usepackage{float}
\usepackage{sectsty}
\usepackage{bm,bbm}
\usepackage[T1]{fontenc}
\usepackage{textcomp}
\usepackage{listings}
\usepackage{caption}
\usepackage{subcaption}
\usepackage{stmaryrd} 
\usepackage{times}

\usepackage[backend=bibtex,sorting=none]{biblatex}
\usepackage{enumitem}
\usepackage[font=footnotesize,format=plain,labelfont=bf]{caption}
\usepackage{multirow}
\usepackage{kotex}
\usepackage{xcolor}
\usepackage{makecell}


\usepackage[colorlinks]{hyperref} 
\hypersetup{ 
    colorlinks=true,       
    linkcolor=red,          
    citecolor=blue,        
    filecolor=magenta,      
    urlcolor=cyan           
}

\newcommand{\eref}[1]{(\ref{#1})} 
\newcommand{\fref}[1]{Fig.~\ref{#1}}
\newcommand{\sref}[1]{Section~\ref{#1}}

\begin{document}

\begin{center}
\textbf{\Large Segment-Based Wall Treatment Model for Heat Transfer Rate in Smoothed Particle Hydrodynamics}

\bigskip

Hyung-Jun Park$^{1}$, Jaekwang Kim$^{2}$, Hyojin Kim$^{3,*}$ \\
\bigskip
\small{
\textit{
$^1$Platform Technology Research Center, LG Chem, 30, Magokjungang 10-ro, Gangseo-gu, Seoul 07796, South Korea \\
$^2$Department of Mechanical and Design Engineering, Hongik University, Sejong 30016, Republic of Korea\\
$^3$Center for Healthcare Robotics, Korea Institute of Science and Technology,
Seoul 02792, Republic of Korea
}

\bigskip
constlearner89@gmail.com, jk12@hongik.ac.kr, hjinkim@kist.re.kr \\
$^*$ Corresponding author 
}
\end{center}

\bigskip

\begin{center}
\textbf{Abstract }\\
\bigskip
\begin{minipage}{0.85\textwidth}
In this study, a smoothed particle hydrodynamics (SPH) model that applies a segment-based boundary treatment is used to simulate natural convection. In a natural convection simulated using an SPH model, the wall boundary treatment is a major issue because accurate heat transfer from boundaries should be calculated. The boundary particle method, which models the boundary by placing multiple layers of particles on and behind the wall boundary, is the most widely used boundary treatment method. Although this method can impose accurate boundary conditions, boundary modeling for complex shapes is challenging and requires excessive computational costs depending on the boundary shape. In this study, we utilize a segment-based boundary treatment method to model the wall boundary and apply this method to the energy conservation equation for the wall heat transfer model. The proposed method solves the problems arising from the use of boundary particles and simultaneously provides accurate heat transfer calculation results for the wall. In various numerical examples, the proposed method is verified through a comparison with available experimental results, SPH results using the boundary particle method, and finite volume method (FVM) results.
\end{minipage}
\end{center}

\newpage

\section{Introduction}

Natural convection is heat transfer mechanism in which density variations of thermal fluids play a significant role in thermal transport.
Natural convection in enclosed systems has been extensively researched owing to its numerous practical applications in various industrial and everyday scenarios, including airflow for a comfortable indoor environment, heat control in data centers through convection, and cooling systems for various industrial facilities~\cite{schaubExperimentalInvestigationHeat2019,kimApplicationArtificialNeural2023}. 
To analyze natural convection, grid-based numerical  methods from a Eulerian perspective such as finite volume method (FVM)~\cite{zhaoHighorderCharacteristicsUpwind2000,abu-nadaEffectsVariableViscosity2009} and finite element method (FEM) ~\cite{kim2021multi,kimNonhomogeneousFlowThixotropic2020,kimAdjointbasedSensitivityAnalysis2023}, are widely used in both industry and academia.
However, these methods frequently require supplementary schemes to manage turbulent flows~\cite{zhaoHighresolutionHighorderUpwind2016,kumarStudySGSClosure2016}, which can develop even under mild external conditions owing to the low-density fluid in natural convection scenarios.
In contrast, smoothed particle hydrodynamics (SPH) has gained attention as a numerical analysis technique for simulating flows with interfaces \cite{monaghanSimulatingFreeSurface1994,baktiComparativeStudyStandard2016,huNumericalSimulationsSloshing2019}.
One of the primary advantages of the SPH scheme when handling interface problems is its particle-based Lagrangian approach, which avoids the need for intricate constraints, such as maintaining high grid quality or grid connectivity, which are commonly required by grid-based methods \cite{liuSmoothedParticleHydrodynamics2003,liuSmoothedParticleHydrodynamics2003a,liuSmoothedParticleHydrodynamics2010a}. 
Moreover, SPH excels in examining fluid dynamics involving free surfaces or multiple phases because interfaces can be  easily tracked by the movement of particles. 
However, when simulating natural convection using SPH, there are still many areas for improvement, including accuracy issues related to boundary condition enforcement at wall boundaries, when compared to its counterparts \cite{liuTreatmentSolidBoundary2012}.

Various formulations have been proposed to simulate natural convection using SPH.
Cleary PW \cite{clearyModellingConfinedMultimaterial1998} applied SPH to energy equations and modeled natural convection based on the Boussinesq approximation.
Szewc et al. \cite{szewcModelingNaturalConvection2011} proposed an SPH model that implemented natural thermal conduction without the Boussinesq approximation.
Building upon pioneering research, Yang and Kong \cite{yangNumericalStudyNatural2019} analyzed the behavior of natural convection based on the Rayleigh and Prandtl numbers using SPH. 
Ng et al. \cite{ngAssessmentSmoothedParticle2020} evaluated two representative boundary particle methods for simulating the heat transfer at walls.
Garoosi et al. \cite{garoosiImprovedHighorderISPH2020} proposed a model for heat transfer analysis in fluid-structure interaction problems by applying a kernel derivative-free model to an incompressible SPH (ISPH). Furthermore, Yang et al. \cite{yangSimulatingNaturalConvection2021} conducted stable simulations of natural convection, even at high Rayleigh numbers, by combining four techniques primarily used for solution stabilization in SPH: kernel gradient correction, particle shifting technique, ${\delta}$-SPH, and asymmetric pressure approximation.
While these studies have contributed to enhancing the accuracy of solutions and proposing various models to simulate natural convection and implement the physical phenomena, they have not addressed the limitations associated with boundaries. 

In SPH, when fluid particles approach a physical solid boundary, they encounter particle deficiency issues. This is known as the boundary truncation problem and is critical because the SPH scheme requires maintenance of a certain number of interacting particles to ensure the accuracy and stability of the solution \cite{bonetVariationalMomentumPreservation1999a,feldmanDynamicRefinementBoundary2007}.
To address these issues in SPH, conventional treatment involves placing multiple layers of boundary particles.  
However, determining the appropriate positions for the boundary particles becomes a challenging task when dealing with complex bounday geometries.
In addition, 
an increase in the number of boundary particles leads to a significant increase in the computational costs \cite{leroyUnifiedSemianalyticalWall2014}.
Hence, we aim to develop a new method that overcomes these limitations while accurately predicting the heat transfer across solid boundaries.

On the other hand, there also exist various methods that do not rely on the use of boundary particles in SPH.
The semi-analytical model proposed by Kulasegaram et al. \cite{kulasegaramVariationalFormulationBased2004} addresses the issue of boundary truncation by applying a normalization factor to the SPH formulation of the fluid particles approaching the boundary region.
Ferrand et al. \cite{ferrandUnifiedSemianalyticalWall2013} proposed an extended unified semi-analytical model that could be applied to complex boundary shapes. 
Mayrhofer et al. \cite{mayrhoferUnifiedSemianalyticalWall2015} extended the unified semi-analytical model into three dimensions, wherea Kostorz and Esmail-Yakas \cite{kostorzSemianalyticalSmoothedparticleHydrodynamics2021} focused on precisely calculating the normalization factor in this model.
More recently, Park et al. \cite{parkDirectImpositionWall2021} introduced a segment-based boundary treatment method that employs line segments in two-dimensions (2D) and triangular segments for three-dimensions (3D) respectively, to model the wall boundary instead of boundary particles.
The use of boundary segments efficiently addresses the particle deficiency at the boundaries, while enhancing the solution accuracy.
Their formulation was primarily applied to the continuity and momentum conservation equations and was used to 
accurately calculate the physical quantities at the boundary.

In this study, by extending the work of Park et al. \cite{parkDirectImpositionWall2021}, 
we propose a wall boundary model for an energy conservation equation that considers heat transfer from the wall boundary. 
In the segment-based boundary treatment method, the boundary truncation problem is addressed by direct compensation of the truncated region through the addition of specific terms called \textit{boundary truncation term} to the original equation.
Whereas previous studies proposed formulations for the boundary truncation term but only for the pressure gradient and viscosity terms in a momentum conservation equation. However, we introduce a boundary truncation term for the energy equation, 
enabling the efficient calculation of heat transfer from wall boundaries composed of segments.
The proposed method addresses issues related to boundary particles and simultaneously delivers precise calculations for heat transfer from the wall.

The rest of the paper is structured as follows: In \sref{sec:GoverningEquations}, we briefly describe the governing equations for non-isothermal buoyant fluid flow. 
A formulation for the boundary truncated terms of the energy equation is proposed in \sref{sec:numerical_method}. The details of the numerical methods are summarized in the same section. 
In \sref{sec:results}, we validate our approach using the results from the experiment and FVM \cite{singhNumericalAnalysisUnsteady2020}. 
Furthermore, the computational performance of the proposed approach is compared with that of a conventional SPH scheme based on the boundary particle method.
The main conclusions are summarized in \sref{sec:conclusion}.

\section{Governing equations}
\label{sec:GoverningEquations}

In this section, we present the governing equations for simulating the non-isothermal buoyant fluid flow. These equations encompass the principles of mass, momentum, and energy conservations:
\begin{equation}
    \frac{{D\rho }}{{Dt}} =  - \rho \nabla  \cdot {\bf{v}},
    \label{eqn:constitutiveEqn}
\end{equation}
\begin{equation}
    \frac{{D{\bf{v}}}}{{Dt}} =  - \frac{1}{\rho }\nabla p + \frac{\mu }{\rho }{\nabla ^2}{\bf{v}} + {{\bf{F}}^B},
    \label{eqn:momentumEqn}
\end{equation}
\begin{equation}
    \frac{{DT}}{{Dt}} =  - \frac{k}{{\rho {C_p}}}{\nabla ^2}T,
    \label{eqn:energyEqn}
\end{equation}
where $\rho $ is fluid density, ${\bf{v}}$ is velocity, $p$ is pressure, $\mu $ is fluid dynamic viscosity, ${{\bf{F}}^B}$ is buoyancy force, $T$ is temperature, $k$ is fluid thermal conductivity, ${C_p}$ is fluid specific heat, and $D/Dt$ denotes the material derivative. 

In this study, the following Boussinesq approximation is used to model the buoyancy force ${{\bf{F}}^B}$ in Eq. \eref{eqn:momentumEqn}:
\begin{equation}
{{\bf{F}}^B} = {\bf{g}}\beta (T - {T_r}),
    \label{eqn:boussinesqApproxi}
\end{equation}
in which ${\bf{g}}$ is the gravitational acceleration, $\beta $ is the fluid thermal expansion coefficient, ${T_r}$ is the reference temperature. When the fluid temperature is higher than the reference temperature ${T_r}$, the fluid experiences an upward buoyancy force. 

The Boussinesq approximation is a model mainly used to simulate buoyancy-driven flows, and this approximation is used in this paper because of its ability to describe the buoyancy effect due to temperature with a simple term \cite{wangModelingHeatTransfer2019}. In other words, once the temperature field is determined using the energy conservation equation in \eqref{eqn:energyEqn}, the buoyancy force associated with the temperature in the next step can be easily calculated.

\begin{figure}
\begin{center}
\includegraphics[width=1\textwidth]{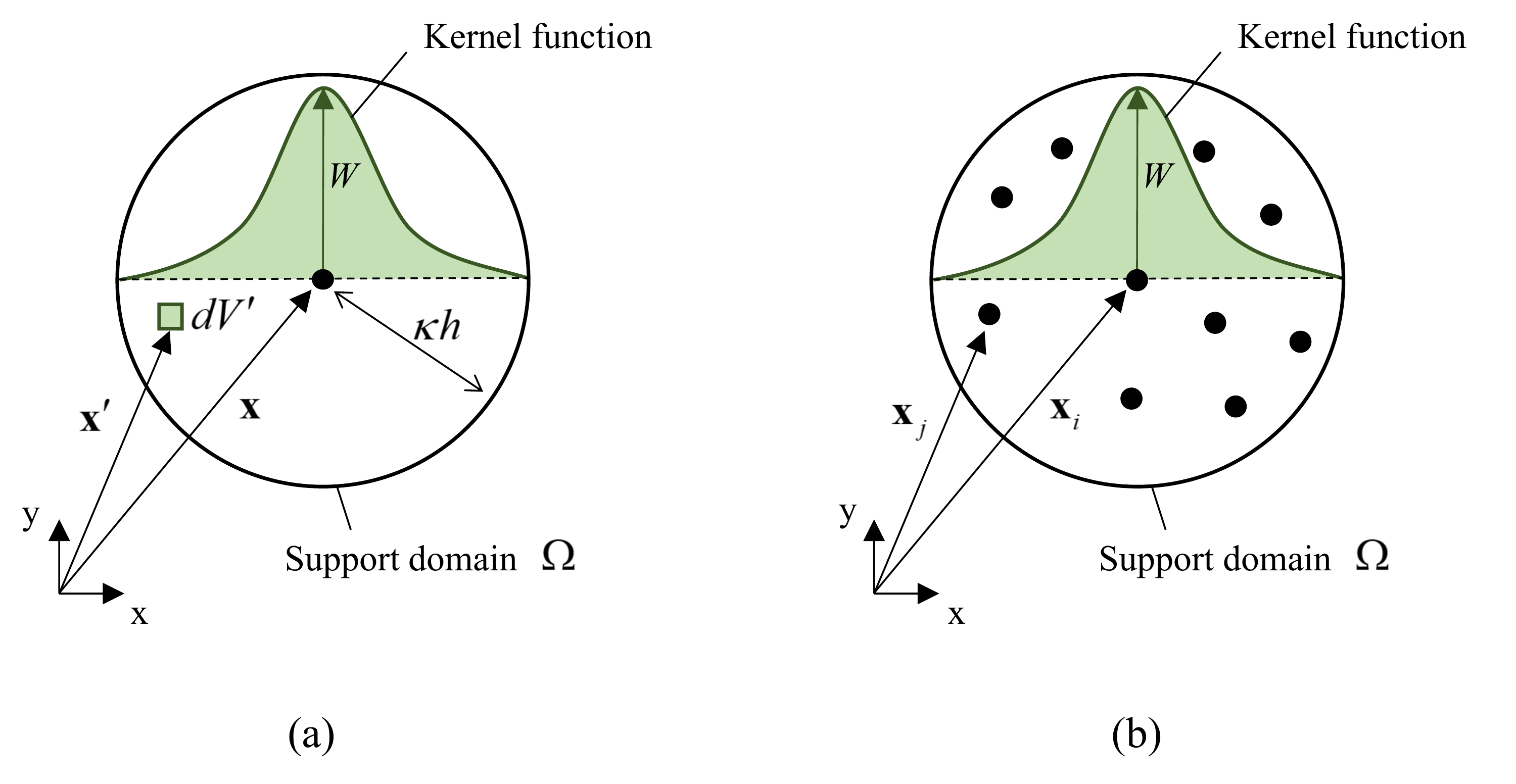}
\end{center}
\caption{Schematic configuration for SPH approximation: (a) kernel approximation and (b) particle approximation.
}\label{fig:SPH_discretization}
\end{figure}

\section{Numerical modeling}
\label{sec:numerical_method}

This section presents a numerical method for discretizing the governing equations. In this sections, the SPH formulation process, which involves the discretization of the spatial domain using particles, is first introduced. Subsequently, the handling of the boundary conditions, wall heat transfer and viscous models, the particle shifting algorithm, and the integration scheme for the time domain are presented.

\subsection{Weakly compressible SPH formulation}
In the problem domain, the SPH method performs a discretization process in two steps. The first step is "kernel approximation", which employes a kernel function, and weight function based on distance, to approximate the field function $f({\bf{x}})$, at position ${\bf{x}}$. Within the valid region of the kernel function, the value of $f({\bf{x}})$ is approximated by integrating the product of the field function $f({\bf{x'}})$ and the kernel function at position ${\bf{x'}}$ within the area. When the kernel approximation is applied to both $f({\bf{x}})$ and its derivatives, the following equations are obtained:

\begin{equation}
    f({\bf{x}})\, \cong \,\int_\Omega  {f({\bf{x'}})\,W({\bf{x}} - {\bf{x'}},h)\,dV'},
    \label{eqn:fx_kernel_approxi}
\end{equation}
\begin{equation}
    \nabla f({\bf{x}})\, \cong \, - \int_\Omega  {f({\bf{x'}})\,\nabla W({\bf{x}} - {\bf{x'}},h)\,dV'},
    \label{eqn:del_fx_kernel_approxi}
\end{equation}
where $W$ represents the kernel function, $h$ is the smoothing length that defines the effective area of the kernel function, $dV'$ is the small volume within the integrated area, and $\Omega $ is the support domain. As shown in \fref{fig:SPH_discretization}(a), the radius of the support domain is determined using $\kappa h$, where $\kappa $ is a parameter dependent on the kernel function. We use the Wendland kernel function and in this case, $\kappa $ is 2 \cite{huIncompressibleMultiphaseSPH2007}.

In the second step "particle approximation", the integral regions of Eqs. \eqref{eqn:fx_kernel_approxi} and \eqref{eqn:del_fx_kernel_approxi} are discretized using a finite number of particles. After placing particle $i$ at the target position ${\bf{x}}$ and particle $j$ at position ${\bf{x'}}$ within the support domain, and substituting the infinitesimal volume $dV'$ with the volume of particle $j$ ($dV' \cong \Delta {V_j} = {m_j}/{\rho _j}$), we obtain the following discretized equations \cite{morrisModelingLowReynolds1997}:

\begin{equation}
    f({{\bf{x}}_i})\, = \sum\limits_{j = 1}^N {f({{\bf{x}}_j})\,W({{\bf{x}}_i} - {{\bf{x}}_j},h)\frac{{{m_j}}}{{{\rho _j}}}},
    \label{eqn:fx_particle_approxi}
\end{equation}
\begin{equation}
    \nabla f({{\bf{x}}_i})\, = \sum\limits_{j = 1}^N {f({{\bf{x}}_j})\,{\nabla _i}W({{\bf{x}}_i} - {{\bf{x}}_j},h)\frac{{{m_j}}}{{{\rho _j}}}},
    \label{eqn:del_fx_particle_approxi}
\end{equation}
in which $m$ is the mass, ${\nabla _i}W$ is a derivative of the kernel function with respect to particle $i$, and the subscripts $i$ and $j$ refer to particles $i$ and $j$, respectively, see \fref{fig:SPH_discretization}(b).

By applying Eqs. \eqref{eqn:fx_particle_approxi} and \eqref{eqn:del_fx_particle_approxi} into the governing equations in Eqs. \eqref{eqn:constitutiveEqn}-\eqref{eqn:energyEqn}, the following SPH formula can be obtained \cite{clearyModellingConfinedMultimaterial1998}:

\begin{equation}
\frac{{D{\rho _i}}}{{Dt}} = {\rho _i}\sum\limits_{j = 1}^N {\frac{{{m_j}}}{{{\rho _j}}}{{\bf{v}}_{ij}}{\nabla _i}W({{\bf{x}}_{ij}},h)}  + {D_i},
\label{eqn:continuity_SPH}
\end{equation}
\begin{equation}
    \frac{{D{{\bf{v}}_i}}}{{Dt}} =  - \sum\limits_{j = 1}^N {{m_j}} \left( {\frac{{{p_i} + {p_j}}}{{{\rho _i}{\rho _j}}}} \right){\nabla _i}W({{\bf{x}}_{ij}},h) + {\bf{F}}_i^V + {\bf{F}}_i^B,
    \label{eqn:momentum_SPH}
\end{equation}
\begin{equation}
    \frac{{D{T_i}}}{{Dt}} =  - \frac{1}{{{\rho _i}{C_p}}}\sum\limits_{j = 1}^N {\frac{{{m_j}}}{{{\rho _j}}}} \left( {\frac{{{k_i}{k_j}}}{{{k_i} + {k_j}}}} \right)\frac{{{{\bf{x}}_{ij}} \cdot {\nabla _i}W({{\bf{x}}_{ij}},h)}}{{{{\left| {{x_{ij}}} \right|}^2}}}({T_i} - {T_j}),
    \label{eqn:energy_SPH}
\end{equation}
in which ${{\bf{v}}_{ij}}$ represents ${{\bf{v}}_i} - {{\bf{v}}_j}$, ${{\bf{x}}_{ij}}$ represents ${{\bf{x}}_i} - {{\bf{x}}_j}$, ${D_i}$ indicates the diffusive term for the density change, and ${{\bf{F}}^V}$ denotes the viscous force term.

The term ${D_i}$ in Eq. \eref{eqn:continuity_SPH} is an addition aimed at stabilizing the density field, and this numerical stabilization approach is referred to as the $\delta $–SPH scheme. In this study, the following equation as proposed by Sun et al. \cite{sunDplusSPHModelSimple2017} is used:

\begin{equation}
{D_i} =  - 2\delta hc\sum\limits_{j = 1}^N {\frac{{{m_j}}}{{{\rho _j}}}{\psi _{ij}}\frac{{{{\bf{x}}_{ij}} \cdot {\nabla _i}W({{\bf{x}}_{ij}},h)}}{{{\bf{x}}_{ij}^2}}} 
\quad \text{with} \quad  
{\psi _{ij}} = ({\rho _j} - {\rho _i}) - \frac{1}{2}\left( {\left\langle {\nabla \rho } \right\rangle _j^L + \left\langle {\nabla \rho } \right\rangle _i^L} \right) \cdot ({{\bf{r}}_j} - {{\bf{x}}_i}),
\end{equation}
where $\delta $ is a parameter that determines the magnitude of density diffusion, $c$ is the numerical speed of sound, and $\left\langle {\nabla \rho } \right\rangle _i^L$ and $\left\langle {\nabla \rho } \right\rangle _j^L$ denote re-normalized density gradient terms, as referred to in \cite{marroneDSPHModelSimulating2011}. Here, $\delta $ is set to 0.1.

In Eq. \eref{eqn:momentum_SPH}, the physical viscosity term ${{\bf{F}}^V}$ is defined as \cite{zhangSPHModelingBubble2015}: 
\begin{equation}
{\bf{F}}_i^V = \sum\limits_{j = 1}^N {\frac{{{m_j}({\mu _i} + {\mu _j})}}{{{\rho _i}{\rho _j}}}\frac{{{{\bf{x}}_{ij}} \cdot {\nabla _i}W({{\bf{x}}_{ij}},h)}}{{{\bf{x}}_{ij}^2 + {{(0.1h)}^2}}}{{\bf{v}}_{ij}}}.
\end{equation}

In weakly compressible SPH (WCSPH), pressure is determined based on the change in density. In this study, the fluid pressure $p$ is calculated using the following equation of state:
\begin{equation}
    {p_i} = {c^2}({\rho _i} - {\rho _0}),
    \label{eqn:EOS}
\end{equation}
in which ${\rho _0}$ represents the initial density of the fluid. The numerical sound speed $c$ is assigned a value that satisfies $c > 10{v_{\max }}$ to impose a weak compressibility condition. ${v_{\max }}$ is the predicted maximum velocity value \cite{ngAssessmentSmoothedParticle2020}, determined by the problem.

\subsection{Boundary treatment}

\begin{figure}
\begin{center}
\includegraphics[width=0.93\textwidth]{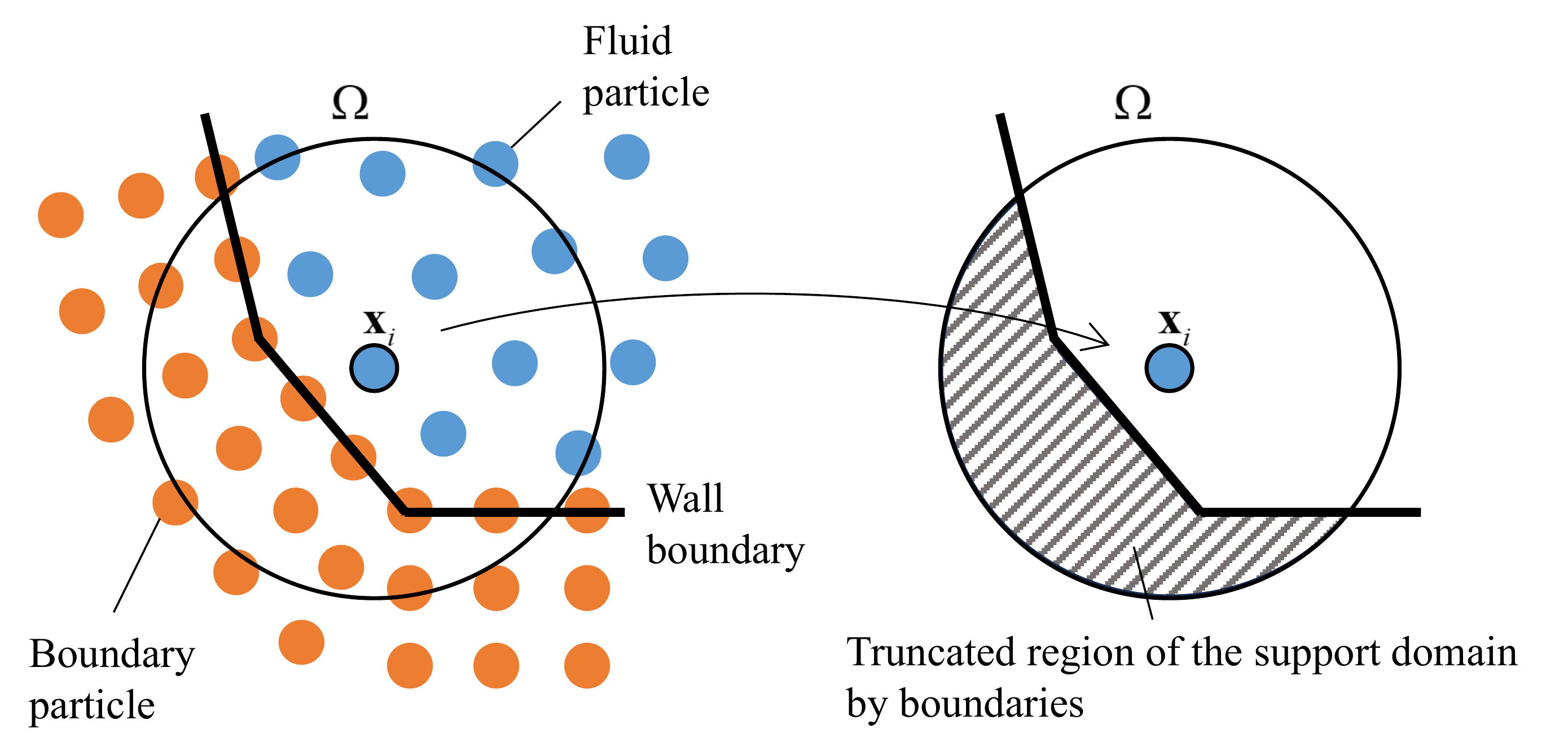}
\end{center}
\caption{Arrangement of the three-layered boundary particles and region of the support domain of the particle $i$, truncated by boundaries.
}\label{fig:truncated_kernel}
\end{figure}

\begin{figure}
\begin{center}
\includegraphics[width=0.46\textwidth]{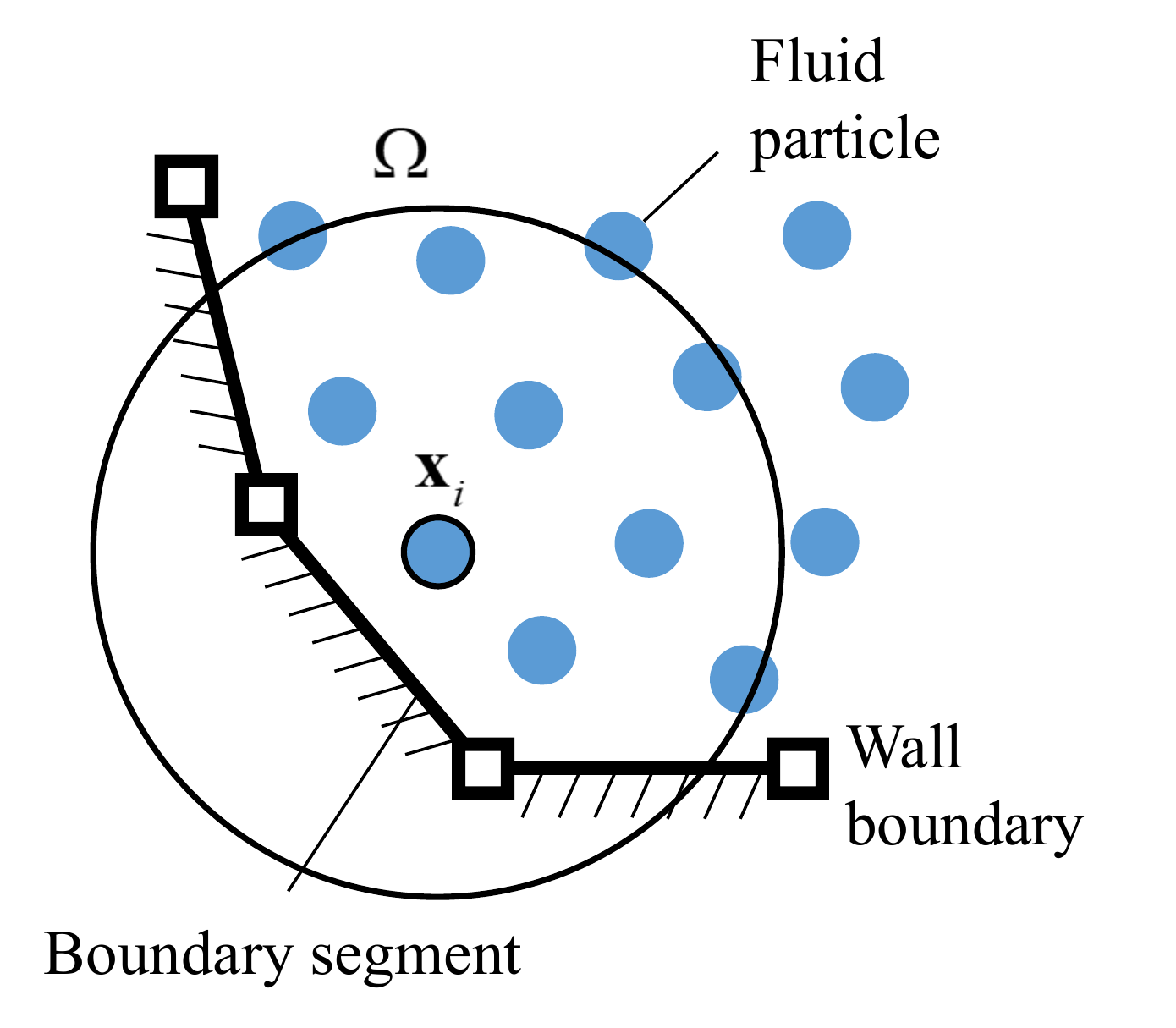}
\end{center}
\caption{Wall boundary modeling using segment-based boundary treatment method.
}\label{fig:segBoundary_schematic}
\end{figure}

Now, we employ the segment-based wall boundary treatment method introduced by Park et al. \cite{parkDirectImpositionWall2021,parkNewSPHFEMCoupling2023}. In traditional SPH boundary treatment methods, multiple layers of particles, referred to as boundary particles, are placed at the boundary, as shown in \fref{fig:truncated_kernel}. These particles serve to restore the support domain of particle $i$, which is truncated near the boundary. Various methodologies, such as the ghost particle method and mirror particle method \cite{liuTreatmentSolidBoundary2012,adamiGeneralizedWallBoundary2012}, exist depending on the placement of the boundary particles or updating of the physical values.

The segment-based boundary treatment method directly addresses the truncated region of the support domain without the placement of boundary particles, see \fref{fig:segBoundary_schematic}. This approach offers several advantages, including avoiding the challenges associated with positioning multiple layers of boundary particles in complex shapes, greater computational efficiency compared with the boundary particle method, and preservation of the original grid shape created by computer aided design (CAD) tools. In other words, boundary modeling can be performed efficiently and easily using CAD tools. 

\begin{figure}
\begin{center}
\includegraphics[width=0.66\textwidth]{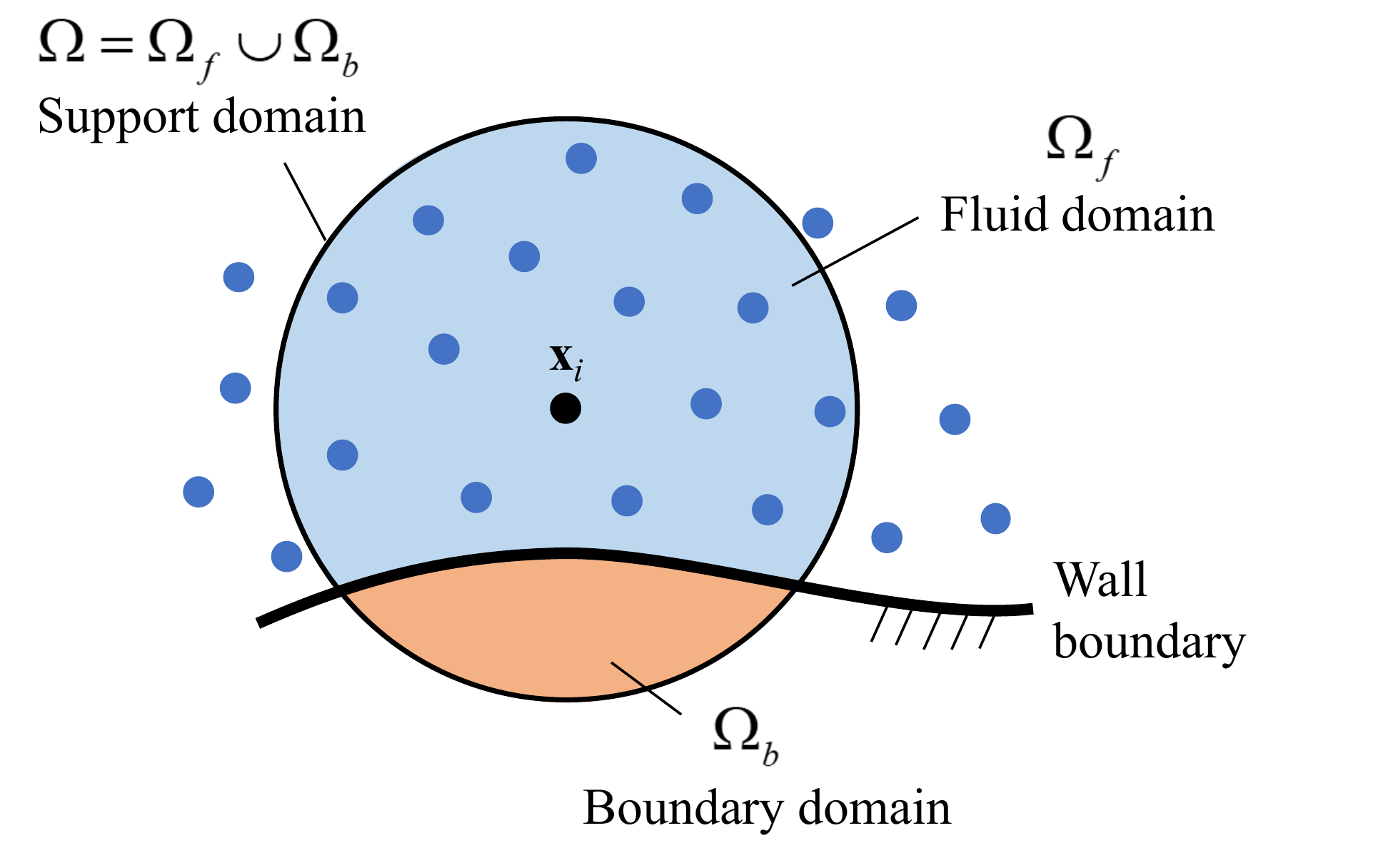}
\end{center}
\caption{Support domain ($\Omega$) divided into fluid ($\Omega _f$) and boundary ($\Omega _b$) domains.
}\label{fig:domain_divided}
\end{figure}

In the segment boundary treatment method, the support domain is divided into the fluid domain ${\Omega _f}$ and the boundary domain ${\Omega _b}$ during the kernel approximation process (\fref{fig:domain_divided}), and Eqs. \eqref{eqn:fx_kernel_approxi} and \eqref{eqn:del_fx_kernel_approxi} can be expressed as follows \cite{parkDirectImpositionWall2021,parkNewSPHFEMCoupling2023}:
\begin{equation}
f({\bf{x}}) \cong \int_{{\Omega _f}} {f({\bf{x'}})WdV'}  + \int_{{\Omega _b}} {f({\bf{x'}})WdV'},
\end{equation}
\begin{equation}
\nabla f({\bf{x}})\, \cong  - \left( {\int_{{\Omega _f}} {f({\bf{x'}})\nabla WdV'}  + \int_{{\Omega _b}} {f({\bf{x'}})\nabla WdV'} } \right),
\label{eqn:del_fx_fluid_boundary}
\end{equation}
in which $W = W({{\bf{x}}_{ij}},h)$. 

\begin{figure}
\begin{center}
\includegraphics[width=0.55\textwidth]{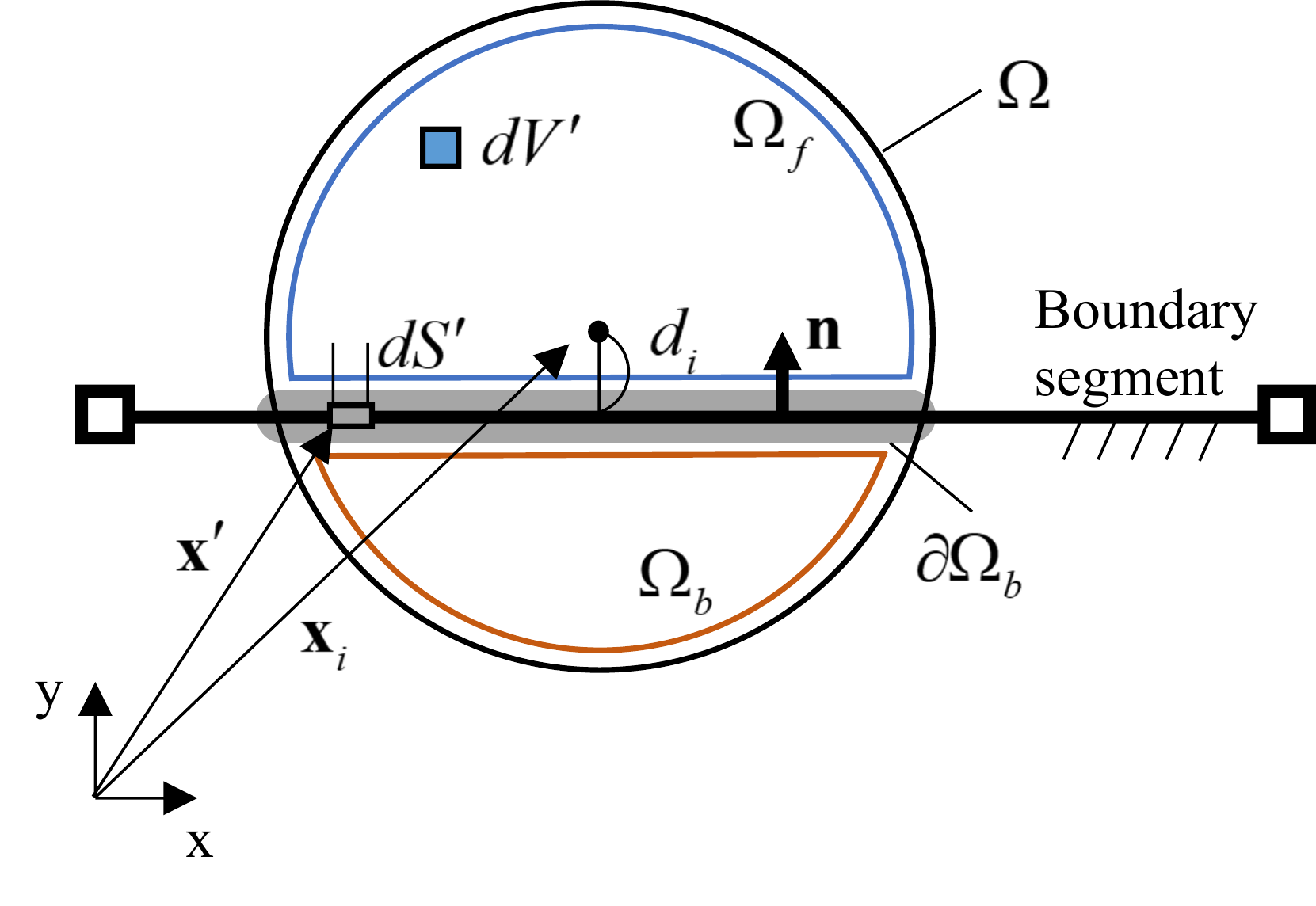}
\end{center}
\caption{Geometric configuration for the segment-based boundary treatment method.
}\label{fig:seg_geometry}
\end{figure}

An examination of Eqs. \eqref{eqn:continuity_SPH}-\eqref{eqn:energy_SPH} shows that these equations primarily comprise the differential terms for the kernel function. Therefore, we utilize only Eq. \eqref{eqn:del_fx_fluid_boundary} to derive the equation for the boundary treatment. By formulating Eq. \eqref{eqn:del_fx_fluid_boundary} in a symmetrical manner and incorporating several inter-domain correlations, following equation can be obtained \cite{parkNewSPHFEMCoupling2023}:
\begin{equation}
\nabla f({{\bf{x}}_i}) = \frac{1}{{\int_{{\Omega _f}}^{} {WdV'} }}\left( {\int_{{\Omega _f}} {\left( {f({\bf{x'}}) - f({{\bf{x}}_i})} \right){\nabla _i}WdV'}  + \int_{\partial {\Omega _b}} {\left( {f({\bf{x'}}) - f({{\bf{x}}_i})} \right)W{\bf{n}}dS'} } \right),
\label{eqn:del_fx_fluid_boundary_2}
\end{equation}
where $\partial {\Omega _b}$ means the cross section where the support domain of particle $i$ is cut by the boundary, $dS'$ is the microscopic area for that face, and ${\bf{n}}$ is a normal vector pointing outward from the face, see \fref{fig:seg_geometry}. Although the derivation process is different, Eq. \eqref{eqn:del_fx_fluid_boundary_2} has the same form as that of a semi-analytical approach, and the denominator on the right hand side of Eq. \eqref{eqn:del_fx_fluid_boundary_2} is called the re-normalization factor, and various studies have been presented to calculate it \cite{leroyUnifiedSemianalyticalWall2014,kostorzSemianalyticalSmoothedparticleHydrodynamics2021}.

In this study, the simplest approach used by Park and Seo \cite{parkNewSPHFEMCoupling2023} is adopted for the calculation of the re-normalization factor. when the particle approximation is applied soley to the fluid domain, including the simplified renormalization factor, Eq. \eqref{eqn:del_fx_fluid_boundary_2} can be expressed as follows:
\begin{equation}
\nabla f({{\bf{x}}_i}) = \frac{1}{{\sum\limits_{j = 1}^N {W\frac{{{m_j}}}{{{\rho _j}}}} }}\left( {\sum\limits_{j = 1}^N {\left( {f({{\bf{x}}_j}) - f({{\bf{x}}_i})} \right){\nabla _i}W\frac{{{m_j}}}{{{\rho _j}}}}  + \int_{\partial {\Omega _b}} {\left( {f({\bf{x'}}) - f({{\bf{x}}_i})} \right)W{\bf{n}}dS'} } \right).
\label{eqn:del_fx_final_form}
\end{equation}

The physical values at the boundary, expressed as $f({\bf{x'}})$ in Eq. \eqref{eqn:del_fx_final_form}, include pressure $p({\bf{x'}})$, velocity ${\bf{v}}({\bf{x'}})$, and temperature $T({\bf{x'}})$. For each boundary value, the following conditions are applied \cite{parkNewSPHFEMCoupling2023}:

\begin{itemize}
    \item 	For pressure $p({\bf{x'}})$
\begin{equation}
p({\bf{x'}}) = {p_i} + \frac{{{\rho _i}}}{{2\Delta t}}({{\bf{v}}_i} - {\bf{v}}({\bf{x'}})) \cdot ({\bf{x'}} - {{\bf{x}}_i}),
\end{equation}

\item For velocity ${\bf{v}}({\bf{x'}})$
\begin{equation}
{\bf{v}}{({\bf{x'}})_{\text{normal}}} = \left( {\frac{{\kappa h}}{{{d_i}}}} \right)({{\bf{v}}_w} \cdot {\bf{n}}) - \left( {\frac{{\kappa h}}{{{d_i}}} - 1} \right)({{\bf{v}}_i} \cdot {\bf{n}}),
\end{equation}
\begin{equation}
{\bf{v}}{({\bf{x'}})_{\text{tangent}}} = \left\{ {\begin{array}{*{20}{c}}{{{\bf{v}}_i} - ({{\bf{v}}_i} \cdot {\bf{n}}){\bf{n}}}&{{\rm{for}}\,{\rm{the}}\,{\rm{free - slip}}\,{\rm{condition,}}}\\{ - \left( {{{\bf{v}}_i} - ({{\bf{v}}_i} \cdot {\bf{n}}){\bf{n}}} \right)}&{{\rm{for}}\,{\rm{the}}\,{\rm{non - slip}}\,{\rm{condition,}}}\end{array}} \right.
\end{equation}

\item For temperature $T({\bf{x'}})$
\begin{equation}
T({\bf{x'}}) = {T_w},
\end{equation}
\end{itemize}
in which $\Delta t$ is the time step size, ${d_i}$ is the distance from the particle $i$ to the boundary plane, ${\bf{v}}{({\bf{x'}})_{normal}}$ and ${\bf{v}}{({\bf{x'}})_{tangent}}$ are the velocities in the normal and tangential directions to the wall, respectively, ${{\bf{v}}_w}$ is the wall velocity, and ${T_w}$ is the wall temperature. Under adiabatic conditions, $T({\bf{x'}}) = {T_i}$, preventing heat transfer from the wall.

The last integral term on the right side of Eq. \eqref{eqn:del_fx_final_form} is directly calculated using the Gaussian integration, and the calculation is performed using the integration algorithm proposed by Ref. \cite{parkDirectImpositionWall2021}.

\subsection{Heat transfer rate and viscous force by wall boundary}
The calculation of the heat transfer rate and viscous force of the wall require discretization involving the Laplacian operator, as shown in Eqs. \eqref{eqn:momentumEqn} and \eqref{eqn:energyEqn}. In this study, the Laplacian operator is discretized using Monaghan's approach \cite{huMultiphaseSPHMethod2006}. The SPH formulation of the Laplacian operator for field functions $f({\bf{x}}{  _i})$ and $g({{\bf{x}}_i})$ is as follows:

\begin{equation}
    \begin{gathered}
    \left( {\frac{1}{\rho }\nabla  \cdot {f_i}\nabla } \right){g_i} \cong \frac{1}{{\sum\limits_{j = 1}^N {W\frac{{{m_j}}}{{{\rho _j}}}} }}\left( {\frac{1}{{{\rho _i}}}\sum\limits_{j = 1}^N {\left( {{f_i} + {f_j}} \right)\left( {\frac{{{g_i} - {g_j}}}{{{{\left| {{{\bf{x}}_{ij}}} \right|}^2}}}} \right)\left( {{{\bf{x}}_{ij}} \cdot {\nabla _i}W} \right)\frac{{{m_j}}}{{{\rho _j}}}} } \right.\\
\left. + \frac{1}{{{\rho _i}}}\int_{\partial {\Omega _b}} {\left( {{f_i} + f'} \right)\left( {\frac{{{g_i} - g'}}{{{{\left| {{{\bf{x}}_i} - {\bf{x'}}} \right|}^2}}}} \right)\left( {({{\bf{x}}_i} - {\bf{x'}}) \cdot {\bf{n}}} \right)WdS'} \right),    
    \label{eqn:Laplacian_SPH}
    \end{gathered}
\end{equation}
where ${f_i} = f({{\bf{x}}_i})$, ${g_i} = g({{\bf{x}}_i})$, $f' = f({\bf{x'}})$, and $g' = g({\bf{x'}})$.

Applying the last term on the right-hand side of Eq. \eqref{eqn:Laplacian_SPH} into the viscosity term of Eq. \eref{eqn:momentumEqn}, we obtain the following wall viscosity term ${\bf{F}}_i^{VW}$:
\begin{equation}
    {\bf{F}}_i^{VW} = \frac{1}{{{\rho _i}\sum\limits_{j = 1}^N {W\frac{{{m_j}}}{{{\rho _j}}}} }}\left( {\int_{\partial {\Omega _b}} {\left( {{\mu _i} + \mu '} \right)\left( {\frac{{{{\bf{v}}_i} - {\bf{v'}}}}{{{{\left| {{{\bf{x}}_i} - {\bf{x'}}} \right|}^2}}}} \right)\left( {({{\bf{x}}_i} - {\bf{x'}}) \cdot {\bf{n}}} \right)WdS'} } \right),
    \label{eqn:wallViscousTerm}
\end{equation}
in which $\mu '$ is assigned the value of ${\mu _i}$.

Similarly, applying the last term on the right-hand side of Eq. \eqref{eqn:Laplacian_SPH} to the energy conversion equation in Eq. \eref{eqn:energyEqn} yields the wall heat transfer rate ${D}T_i^W{/Dt}$:
\begin{equation}    
    \frac{{DT_i^W}}{{Dt}} = \frac{1}{{{\rho _i}{C_p}\sum\limits_{j = 1}^N {W\frac{{{m_j}}}{{{\rho _j}}}} }}\left( {\int_{\partial {\Omega _b}} {\left( {{k_i} + k'} \right)\left( {\frac{{{T_i} - T'}}{{{{\left| {{{\bf{x}}_i} - {\bf{x'}}} \right|}^2}}}} \right)\left( {({{\bf{x}}_i} - {\bf{x'}}) \cdot {\bf{n}}} \right)WdS'} } \right),
    \label{eqn:wallHeatTransferRate}
\end{equation}
in which $k'$ is assigned the value of ${k_i}$.

In Eq. \eqref{eqn:Laplacian_SPH}, the arithmetic mean $({f_i} + f')/2$ for the field function $f$ can be replaced with the harmonic mean $2{f_i}f'/({f_i} + f')$ \cite{huIncompressibleMultiphaseSPH2007,huMultiphaseSPHMethod2006}.

\begin{figure}[t]
\begin{center}
\includegraphics[width=0.61\textwidth]{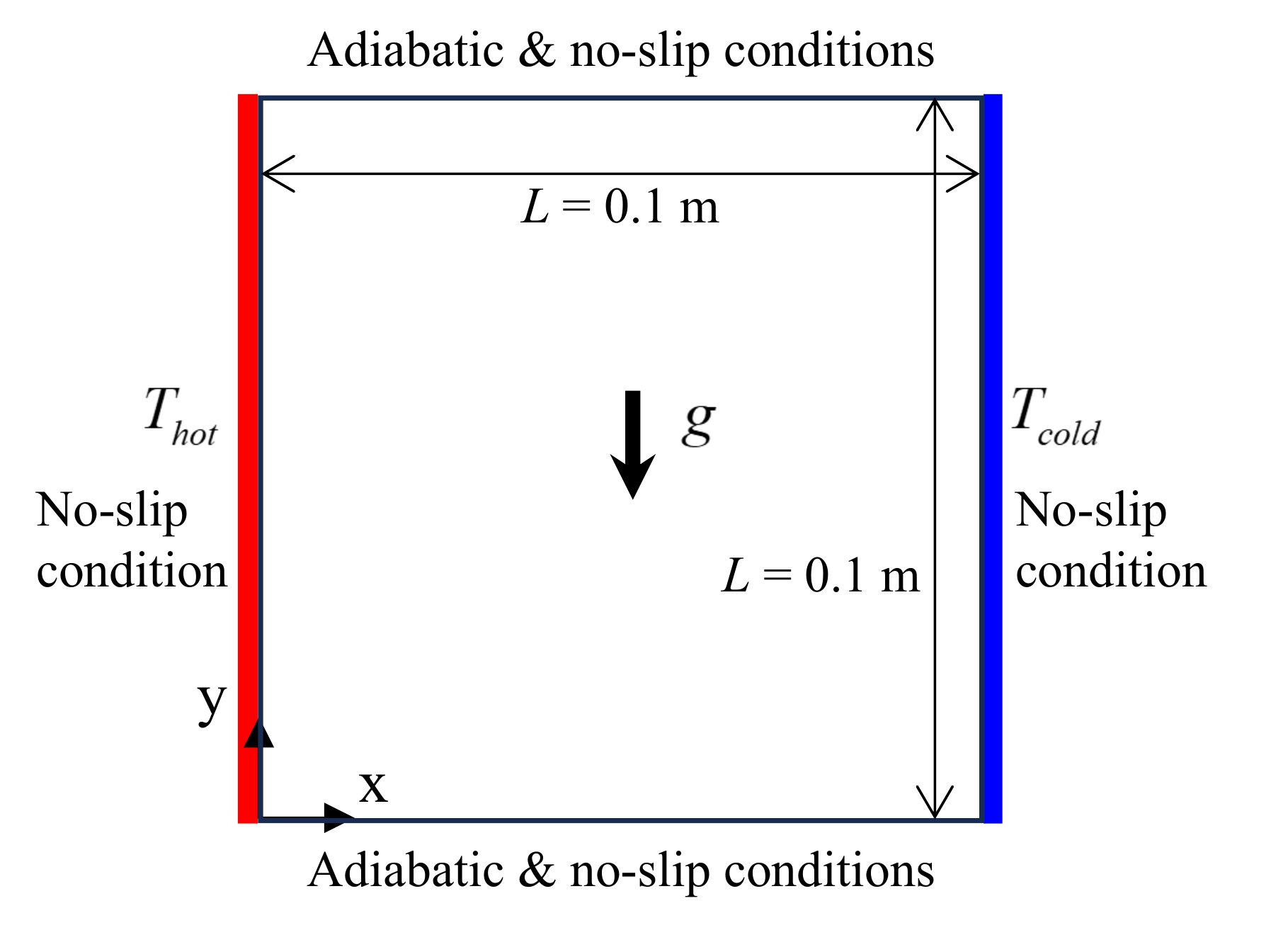}
\end{center}
\caption{Domain description for natural convection in a square cavity problem.
}\label{fig:squareCavity_domain}
\end{figure}

\subsection{Particle shifting algorithm}

In the SPH simulations, local particle disordering can occur as the time step progresses. This disordering of particles causes a particle clustering phenomenon called tensile instability and yields non-physical analysis results \cite{lindIncompressibleSmoothedParticle2012}. To alleviate this disordering, the particle shifting technique (PST) is commonly used in the field of SPH. In this study, we use the approach proposed by Sun et al. \cite{sunDplusSPHModelSimple2017} as follows:
\begin{equation}
    \delta {r_i} =  - 2(\Delta t)\left| {{{\bf{v}}_{\max }}} \right|(\kappa h)\sum\limits_{j = 1}^N {\left[ {1 + R{{\left( {\frac{W}{{W(\Delta x)}}} \right)}^n}} \right]{\nabla _i}W\frac{{{m_j}}}{{({\rho _i} + {\rho _j})}}},
    \label{eqn:particleShiftingEqn}
\end{equation}
where $\Delta x$ represents the initial interparticle spacing, and $R$ and $n$ are problem-dependent values, which were set to 2 and 0.4, respectively, in this study.

Once the change in velocity at each time step is calculated using the momentum conservation equation, it cab be used to compute the velocity value for the next step. This, in turn, determines the position value for the next step. Eq. \eqref{eqn:particleShiftingEqn} is then added to the determined position value to alleviate the particle disorder.

Considering boundary truncation near the boundary, Eq. \eqref{eqn:particleShiftingEqn} can be written as \cite{parkNewSPHFEMCoupling2023}:
\begin{equation}
    \begin{array}{l}\delta {r_i} =  - 2(\Delta t){v_{\max }}(\kappa h)\left( {\sum\limits_j^N {\left[ {1 + R{{\left( {\frac{W}{{W(\Delta x)}}} \right)}^n}} \right]{\nabla _i}W\frac{{{m_j}}}{{({\rho _i} + {\rho _j})}}} } \right.\\\,\,\,\,\,\,\,\,\,\,\, + \left. {\frac{1}{2}\int_{{\Omega _{\,b}}} {\left[ {1 + R{{\left( {\frac{{W({{\bf{x}}_i} - {\bf{x'}})}}{{W(\Delta x)}}} \right)}^n}} \right]{\nabla _i}WdV'} } \right)\end{array}.
    \label{eqn:particleShiftingEqn_proposed}
\end{equation}

\begin{figure}[t]
\begin{center}
\includegraphics[width=1.0\textwidth]{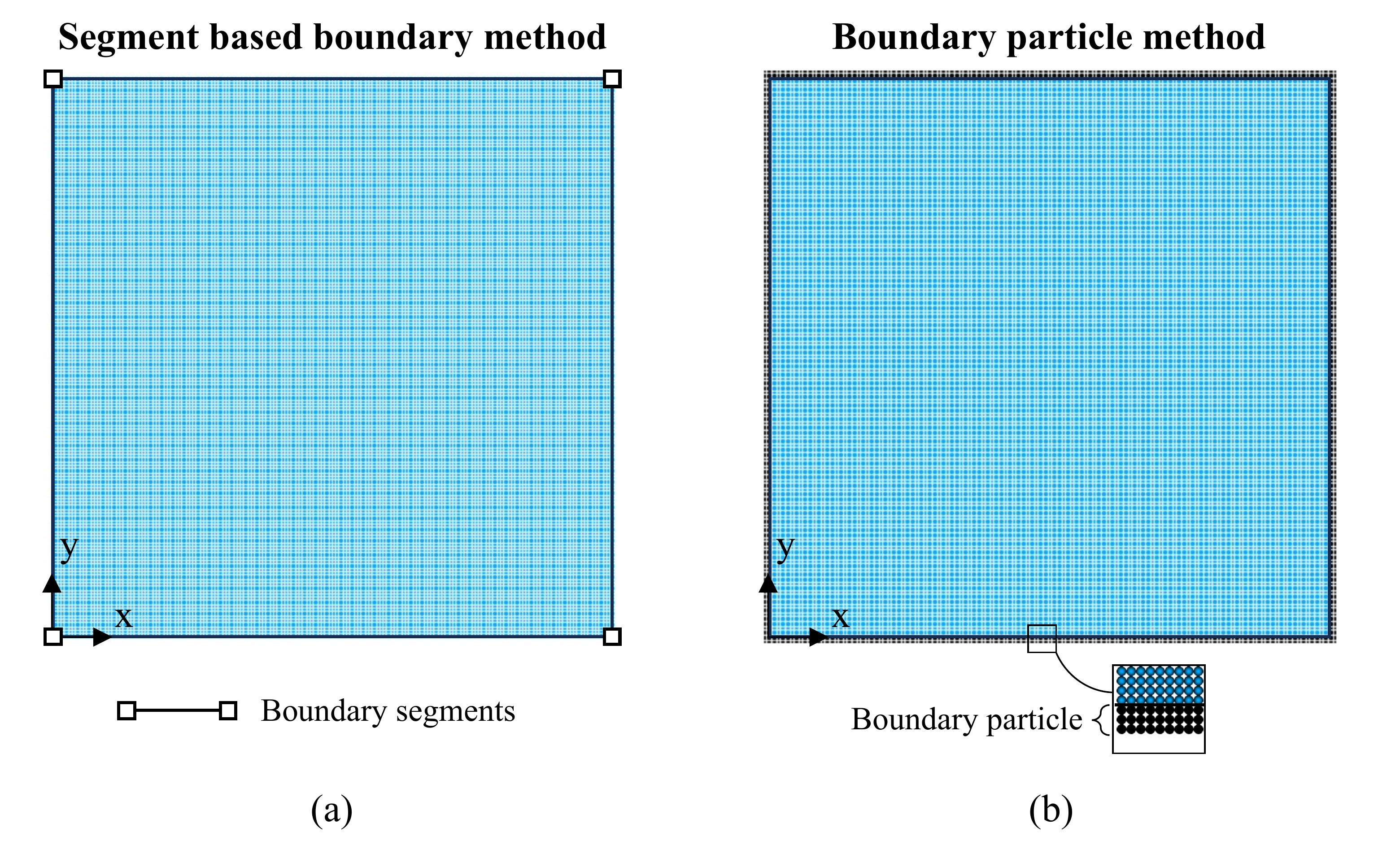}
\end{center}
\caption{Particle discretization domain for natural convection in a square cavity problem using the (a) segment-based boundary treatment method (proposed) and (b) boundary particle method.
}\label{fig:squareCavity_seg_BPM}
\end{figure}

\subsection{Time integration scheme}

In SPH, changes in physical quantities over time are explicitly calculated, and a predictor-corrector time integration scheme is used in this study \cite{monaghanProblemPenetrationParticle1989}. The position, velocity, and density of the particle are updated in the prediction step as follows:
\begin{equation}
    {\bf{x}}_i^{t + \Delta t/2} = {\bf{x}}_i^t + \frac{{\Delta t}}{2}{\bf{v}}_i^{t + \Delta t/2},
    \label{eqn:predict_position}
\end{equation}
\begin{equation}
    {\bf{v}}_i^{t + \Delta t/2} = {\bf{v}}_i^t + \frac{{\Delta t}}{2}{\left( {\frac{{D{{\bf{v}}_i}}}{{Dt}}} \right)^t},
    \label{eqn:predict_velocity}
\end{equation}
\begin{equation}
    \rho _i^{t + \Delta t/2} = \rho _i^t + \frac{{\Delta t}}{2}{\left( {\frac{{D{\rho _i}}}{{Dt}}} \right)^t},
    \label{eqn:predict_rho}
\end{equation}
in which the superscript $t + \Delta t/2$ represents the value predicted in the prediction step. In the correction step, the governing equation is solved based on the physical quantity predicted in the prediction step and the value in the next step is determined as follows:
\begin{equation}
    {\bf{x}}_i^{t + \Delta t} = {\bf{x}}_i^t + \Delta t{\bf{v}}_i^{t + \Delta t},
    \label{eqn:correct_position}
\end{equation}
\begin{equation}
    {\bf{v}}_i^{t + \Delta t} = {\bf{v}}_i^t + \Delta t{\left( {\frac{{D{{\bf{v}}_i}}}{{Dt}}} \right)^{t + \Delta t/2}},
    \label{eqn:correct_velocity}
\end{equation}
\begin{equation}
    \rho _i^{t + \Delta t} = \rho _i^t + \Delta t{\left( {\frac{{D{\rho _i}}}{{Dt}}} \right)^{t + \Delta t/2}},
    \label{eqn:correct_rho}
\end{equation}
in which the superscript $t + \Delta t$ represents the next step.

For numerical stability, the time step size $\Delta t$ should be determined using the Courant-Friedrichs-Lewy (CFL) condition. In SPH, the CFL condition is given by \cite{demouraCourantFriedrichsLewy2013}:
\begin{equation}
    \Delta t \le 0.5\frac{h}{{c + \left| {{{\bf{v}}_{\max }}} \right|}}.
    \label{eqn:CFL_condition}
\end{equation}

For all numerical examples in this study, simulations are performed by fixing $\Delta t$ to satisfy Eq. \eqref{eqn:CFL_condition}.

\section{Results}
\label{sec:results}

In this section, we present three numerical examples to validate the proposed method: natural convection in a square cavity, natural convection in a horizontal concentric annulus, and a complex-shaped cavity. We assess the accuracy and robustness of the proposed method using two benchmark problems that are commonly used to validate numerical simulations of natural convection. Moreover, we demonstrate the versatility of our method by applying it to an example with complex boundary shapes. For all the examples, we verify the proposed method through SPH using boundary particles and FVM, a representative numerical method. For FVM, we use OpenFOAM, a widely used open-source tool. For the boundary particle method, the model proposed in Ref.~\cite{liuTreatmentSolidBoundary2012} is used, and three-layered boundary particles are used for all examples.

\begin{table}[]
\caption{For the SPH results using Boundary particle method, calculation time and particle number ratio compared to the proposed method, and number of boundary segments used in the proposed method.}
\begin{tabular}{@{}llll@{}}
\toprule
Problem domain              & Square   cavity & Horizontal   concentric annulus & Complex   shaped cavity \\ \midrule
Computational time   ratio  & 0.9294          & 0.85                            & 0.8609                  \\
Number of particles   ratio & 0.9429          & 0.9226                          & 0.9301                  \\
Number of segments          & 4               & 60                              & 10                      \\ \bottomrule
\end{tabular}
\label{tab:results_comparison}
\end{table}

Table \ref{tab:results_comparison} shows the ratio between the calculation times of the proposed method and the boundary particle method for each example. In all three numerical examples, the calculation times of the proposed method are approximately faster 8-15\% faster. This shows slightly reduction in computational efficiency compared with the research results using the existing segment-based boundary treatment method \cite{parkDirectImpositionWall2021}, and 
can be attributed to the ratio of the total number of particles used to model the domain. All three numerical examples performed in this study have problem domains in which the fluid particles are fully filled within boundaries. That is, the number of fluid particles is relatively greater than that of boundary particles. In other words, because the total number of particles used in the boundary particle method and the proposed method are not significantly different, the difference in computational efficiency appears to be smaller than reported in previous studies.



\begin{table}[t]
\caption{OpenFOAM setting information. Note that All setting names use the OpenFOAM setting names for the reader's convenience.}
\begin{tabular}{@{}lllll@{}}
\toprule
\multicolumn{2}{l}{\makecell{Problem domain}}                           & \makecell{Square \\ cavity}    & \makecell{Horizontal concentric\\ annulus}   & 
\makecell{Complex shaped \\ cavity }
\\ \midrule
\multicolumn{2}{l}{Mean grid size (m)}         & 0.00094          & 0.00052                         & 0.0047                  \\
\multicolumn{2}{l}{deltaT (s)}                 & 5$\times {10^{ - 4}}$           & 2.5$\times {10^{ - 4}}$                        & 0.02                    \\ \midrule
\multicolumn{2}{l}{Turbulence model}           & \multicolumn{3}{l}{kOmegaSST}                                                \\
\multicolumn{2}{l}{Wall treatment}             & \multicolumn{3}{l}{kqRWallFunction   / omegaWallFunction / nutUWallFunction} \\
\multicolumn{2}{l}{Pressure-velocity coupling} & \multicolumn{3}{l}{PIMPLE}                                                   \\ \midrule
{\hspace{3pt} fvSchemes}         & ddtSchemes                 & \multicolumn{3}{l}{Euler}                                                    \\
                  & gradSchemes                & \multicolumn{3}{l}{Gaus   linear}                                            \\
                  & divSchemes                 & \multicolumn{3}{l}{Bounded   Gauss limitedLinear}                            \\
                  & laplacianSchemes           & \multicolumn{3}{l}{Gauss   linear orthogonal}                                \\
                  & interpolationSchemes       & \multicolumn{3}{l}{Linear}                                                   \\
                  & snGradSchemes              & \multicolumn{3}{l}{orthogonal}                                               \\ \midrule
{\hspace{3pt} thermoType}        & Type                       & \multicolumn{3}{l}{heRhoThermo}                                              \\
                  & mixture                    & \multicolumn{3}{l}{pureMixture}                                              \\
                  & transport                  & \multicolumn{3}{l}{const}                                                    \\
                  & Thermo                     & \multicolumn{3}{l}{hConst}                                                   \\
                  & equationOfState            & \multicolumn{3}{l}{Boussinesq}                                               \\
                  & Specie                     & \multicolumn{3}{l}{specie}                                                   \\
                  & Energy                     & \multicolumn{3}{l}{sensibleEnthalpy}                                         \\ \bottomrule
\end{tabular}
\label{tab:OpenFOAM_settings}
\end{table}

\subsection{Natural convection in a square cavity}

\begin{figure}
\begin{center}
\includegraphics[width=1.0\textwidth]{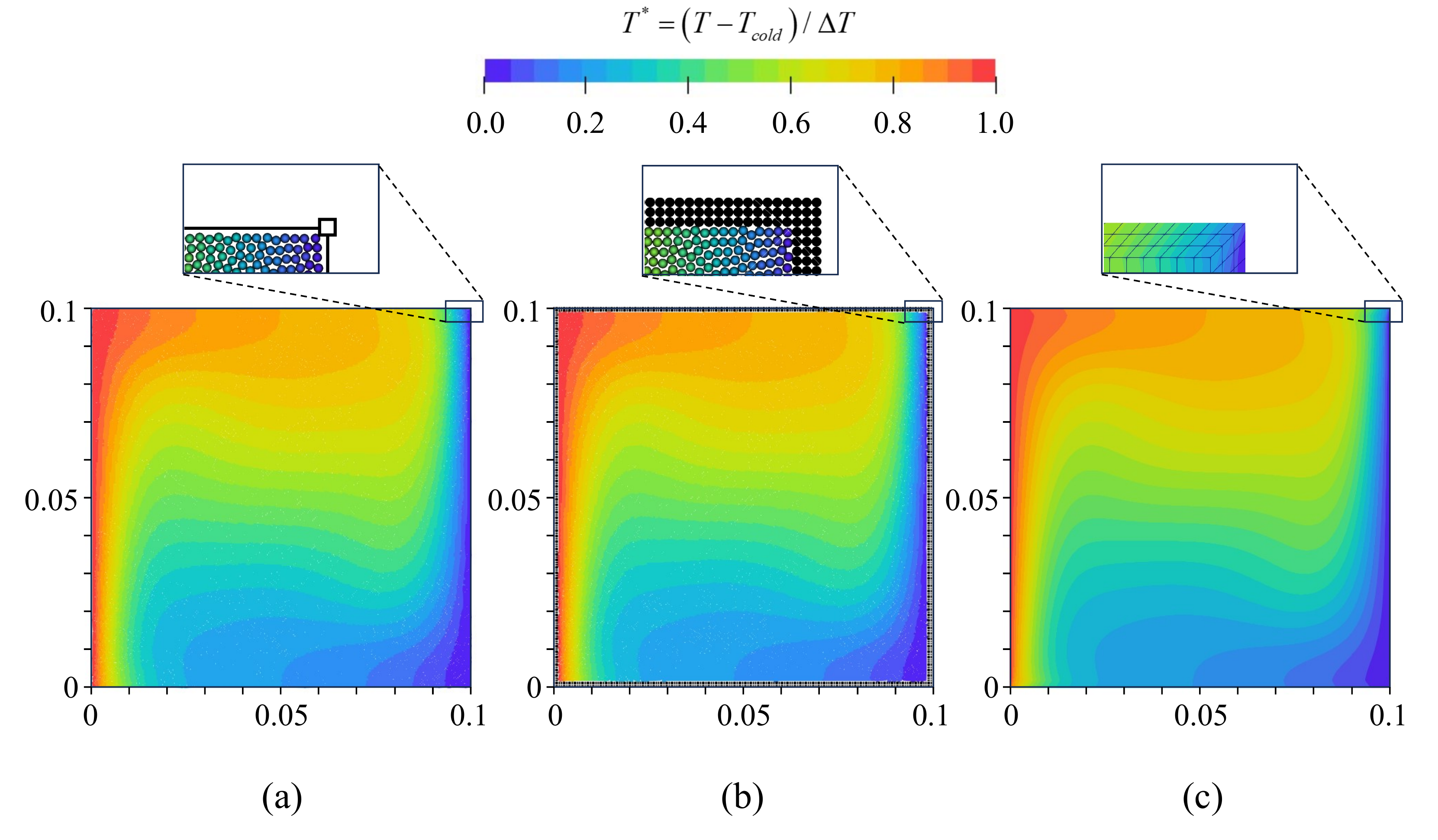}
\end{center}
\caption{Comparison of the temperature contours for natural convection in a square cavity problem among three methods: (a) SPH with segment-based boundary treatment method (proposed), (b) SPH with boundary particle method, and (c) FVM.
}\label{fig:squareCavity_contour}
\end{figure}

\begin{figure}
\begin{center}
\includegraphics[width=1.0\textwidth]{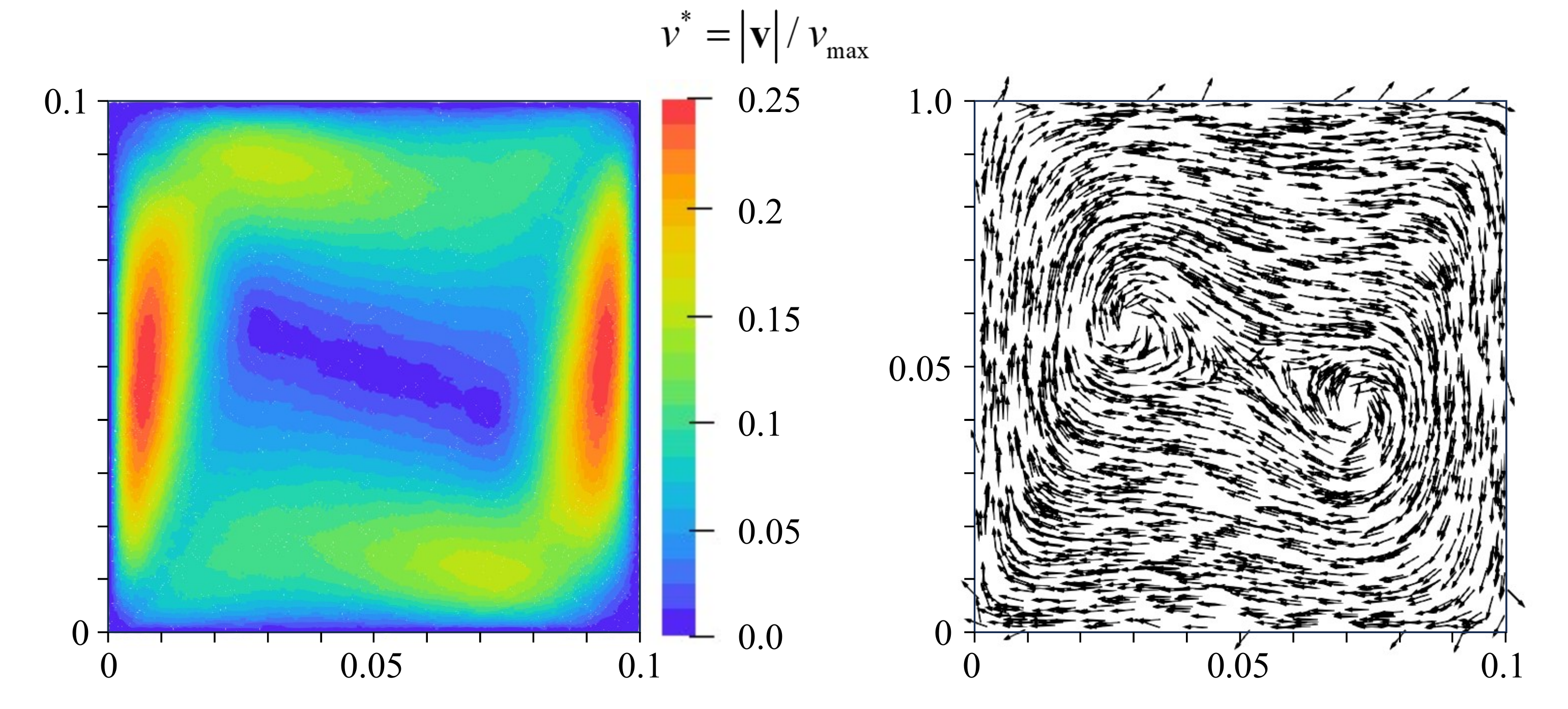}
\end{center}
\caption{Velocity field distribution from the results of the proposed method for natural convection in a square cavity problem: (a) velocity magnitude contour and (b) velocity vector distribution. Note that the velocity vector arrow size is not scaled to its magnitude.
}\label{fig:squareCavity_contour_vel}
\end{figure}

We consider the natural convection in a square cavity as the first numerical example. As shown in \fref{fig:squareCavity_domain}, this is an example in which buoyant flow occurs inside a rectangular box owing to the different temperature distributions on both side walls. This example has a very simple geometry, making it easy to define dimensionless coefficients such as the Prandtl and Rayleigh numbers associated with thermo–fluid dynamics. Because of these characteristics, this benchmark problem has been extensively tested as a validation case in various numerical methods.

The length $L$ of one side of the square box is set to 0.1 m. The temperature of the left wall ${T_{hot}}$ is 300.53487 K, the temperature of the right wall ${T_{cold}}$ is 299.496513 K, and the temperature difference $\Delta T$ ($ = {T_{hot}} - {T_{cold}}$) between the two walls is 1.006974 K. The initial density of the internal fluid ${\rho _0}$ is 1.2 kg/m$^3$, dynamic viscosity $\mu $ is 1.846$ \times {10^{ - 5}}$ Pa s, thermal expansion coefficient $\beta $ is 0.0034 K$^{ - 1}$, specific heat capacity ${C_p}$ is 1000 J/(kg K), and thermal conductivity $k$ is 0.0262 W/(m k), the initial temperature distribution ${T_0}$ is 300 K, and the gravitational acceleration $g$ is 9.81 m/s$^{ - 2}$. In this case, the dimensionless Rayleigh number $Ra$ ($ = {\rho _0}^2{C_p}g\beta \Delta T{L^3}/(k\mu )$), which represents the thermal behavior characteristics of the fluid, is 100,000, and the Prandtl number $Pr$ ($ = {C_p}\mu /k$) is 0.7046.

In the SPH simulation, the initial particle spacing $\Delta x$ is set to confirm the convergence according to particle resolution through three cases of 0.002, 0.001, and 0.0005 m, and the time step size for each case, chosen to satisfies Eq. \eqref{eqn:CFL_condition}, are 2$ \times {10^{ - 3}}$, 1$ \times {10^{ - 3}}$, and 5$ \times {10^{ - 4}}$ s, respectively. ${v_{\max }}$ to calculate the numerical speed of sound $c$ in Eq. \eref{eqn:EOS} is calculated as ${v_{\max }} = \sqrt {g\beta \Delta TL} $. Figs. \ref{fig:squareCavity_seg_BPM}(a) and (b) show the particle discretization domain for the proposed method and boundary particle method, respectively, where $\Delta x$= 0.0005 m. For the slip condition, a no-slip boundary condition is imposed to all wall boundaries.

In the FVM simulation, 11200 grids are used to model the problem domain, and the time step size is 5$ \times {10^{ - 4}}$ s. It is simulated using an incompressible transient solver, the turbulence model is k-omega, and the state equation is Boussinesq approximation. The detailed setting information is provided in Table \ref{tab:OpenFOAM_settings}.

\begin{figure}
\begin{center}
\includegraphics[width=1.0\textwidth]{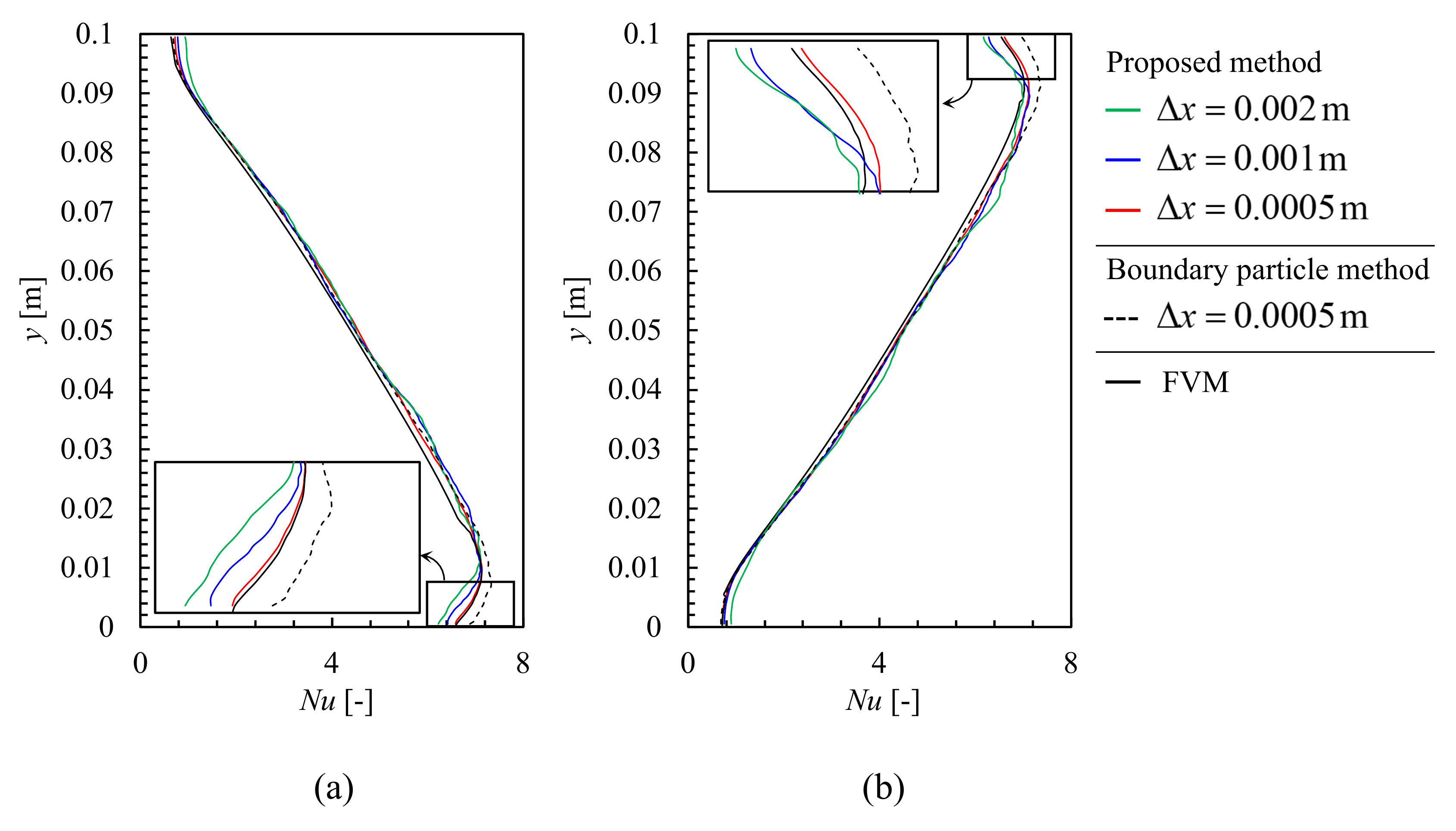}
\end{center}
\caption{$Nu$ profile comparison among three methods (the proposed method, boundary particle method, and FVM) and convergence test at $dx$=0.002, 0.001, and 0.0005 m: (a) $Nu$ profile along a hot wall and (b) $Nu$ profile along a cold wall.
}\label{fig:squareCavity_results}
\end{figure}

\fref{fig:squareCavity_contour}(a) shows the temperature contour for the $\Delta x$= 0.0005 m case at $t$ = 50 s. On the left, at the hot wall, a thick layer of fluid with higher temperature forms as the fluid temperature increases with height. On the right, at the cold wall, a thick layer of fluid with lower temperature forms as the fluid temperature decreases with downward movement. As shown in \fref{fig:squareCavity_contour_vel}, this temperature distribution induces a buoyancy force in the fluid, resulting in counterclockwise circulation in the outer fluid (closer to the wall). On the other hand, vortices rotating in opposite directions are observed on both sides of the center. Figs. \ref{fig:squareCavity_contour}(b) and (c) show the results of SPH using boundary particles and FVM, respectively, demonstrating a qualitative alignment with the outcomes of the proposed method.

To verify the accuracy of the proposed method through results of each method, the distribution of Nusselt number $Nu$ ($ =  - \frac{{dT}}{{dx}}\frac{L}{{\Delta T}}$) is compared quantitatively. $Nu$ is the ratio of the heat transfer by conduction to that by convection at the boundary surface. The distributions of $Nu$ on both sides of the wall are compared. Figs. \ref{fig:squareCavity_results}(a) and (b) show the comparisons of the three methods for the hot and cold walls, respectively. The quantitative results of the proposed method tend to be in good agreement with the results of SPH using the boundary particles and FVM. In addition, the convergence of the proposed method is confirmed to be closer to the FVM results, as the particle resolution becomes finer.

\begin{figure}
\begin{center}
\includegraphics[width=0.73\textwidth]{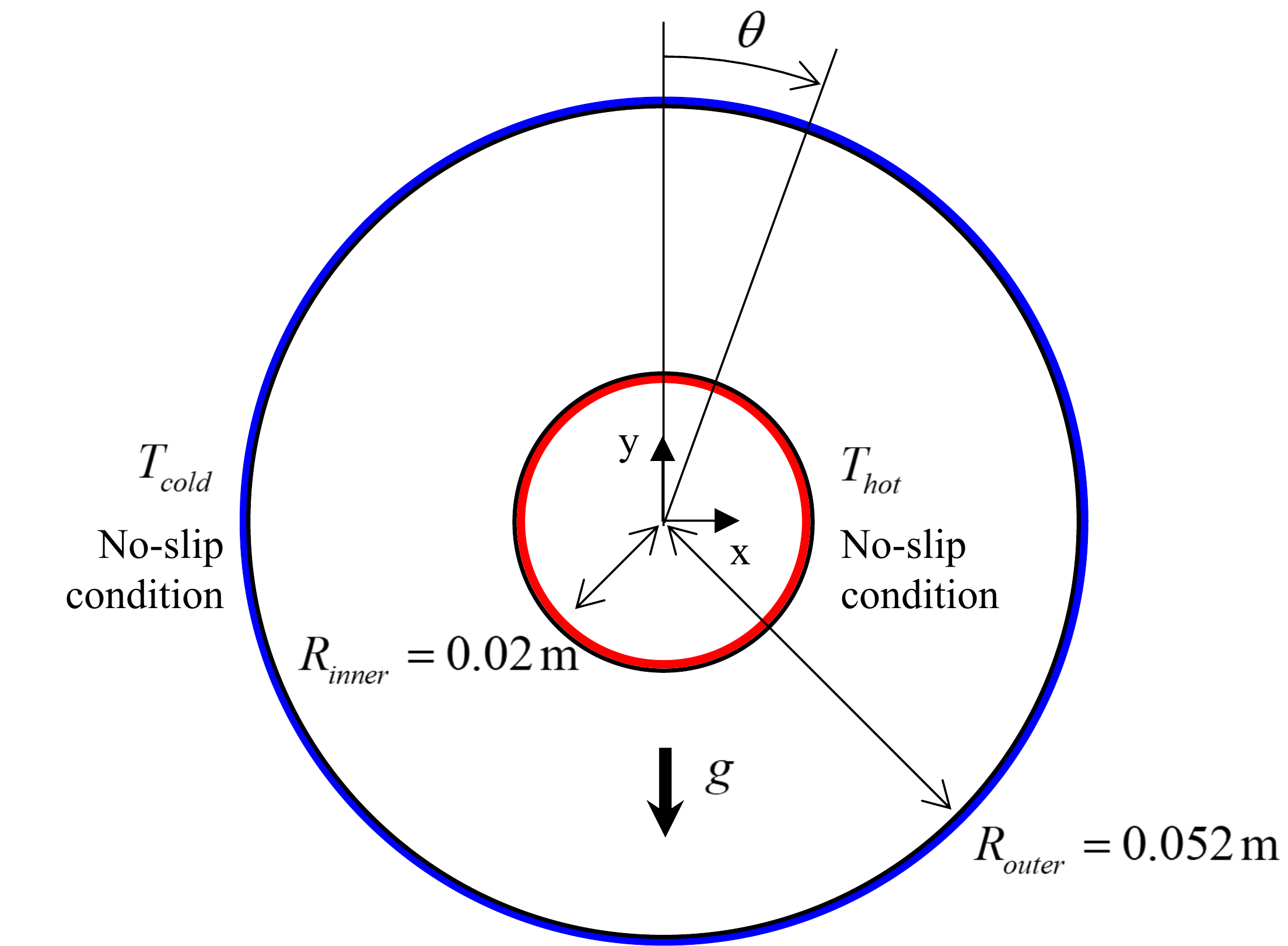}
\end{center}
\caption{Domain description for natural convection in a horizontal concentric annulus problem.
}\label{fig:conCylinder_domain}
\end{figure}

\begin{figure}
\begin{center}
\includegraphics[width=0.95\textwidth]{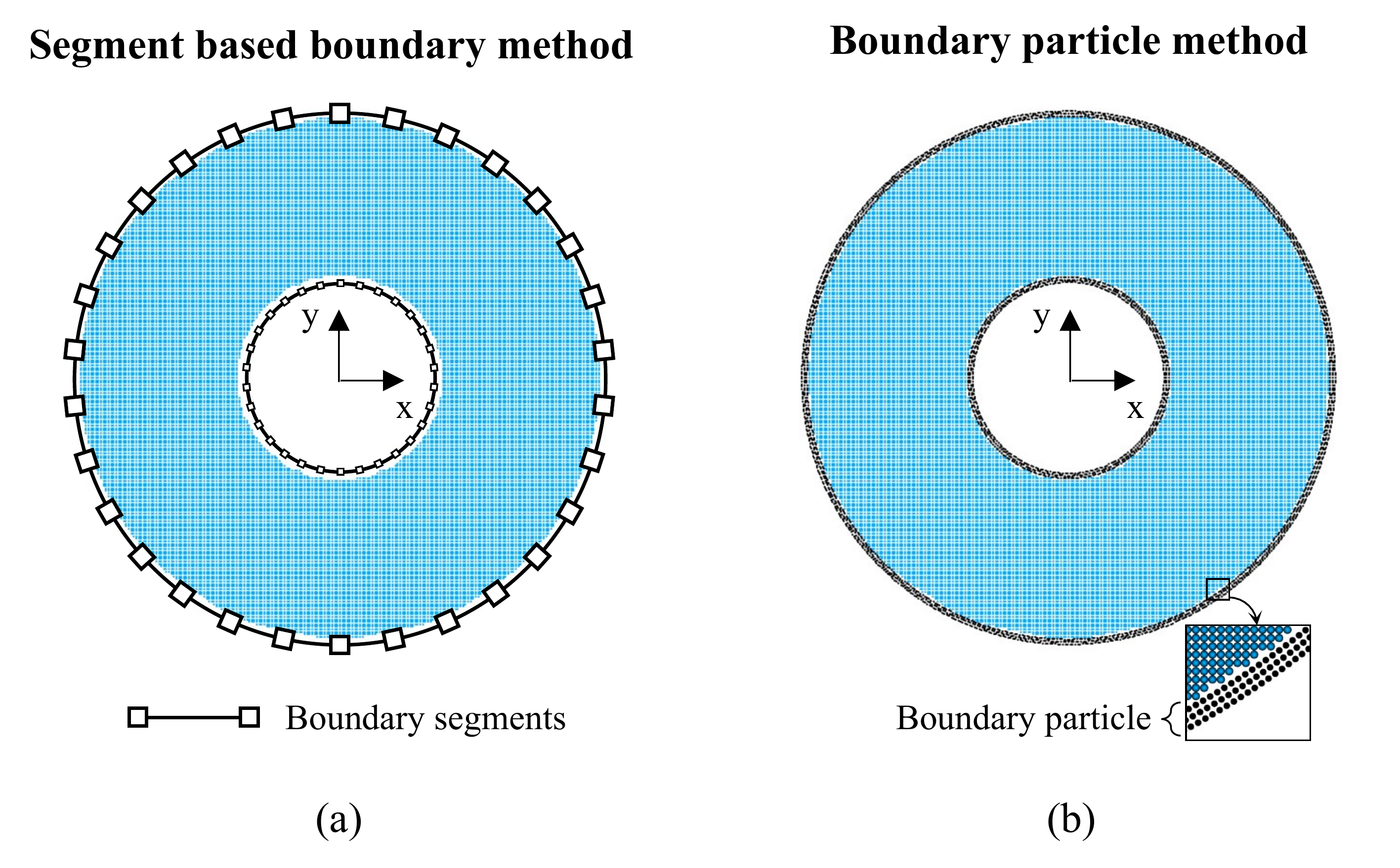}
\end{center}
\caption{Particle discretization domain for natural convection in a horizontal concentric annulus problem using (a) segment-based boundary treatment method (proposed) and (b) boundary particle method.
}\label{fig:conCylinder_seg_BPM}
\end{figure}

\subsection{Natural convection in a horizontal concentric annulus}

We consider the natural convection in a horizontal concentric annulus, shown in \fref{fig:conCylinder_domain}, as the second numerical example. Natural convection occurs owing to the temperature difference between the inner and the outer cylinders. This example allows us verify the numerical analysis results against the experiments performed by \cite{kuehnExperimentalStudyNatural1978}. That is, the proposed method is validated by comparison with experimental results as well as other numerical approaches such as FVM and SPH with boundary particles.

The radius ${R_{inner}}$ and temperature ${T_{hot}}$ of the inner circle are 0.02 m and 323.664 K, respectively. the radius ${R_{outer}}$ and temperature ${T_{cold}}$ of the outer circle are 0.052 m and 300 K, respectively. The temperature difference $\Delta T$ between the inner and outer circles is 23.664 K. The initial density of the internal fluid ${\rho _0}$ is 1.096 kg/m$^3$, dynamic viscosity $\mu $ is 2.0$ \times {10^{ - 5}}$ Pa s, thermal expansion coefficient $\beta $ is 0.003 K$^{ - 1}$, specific heat capacity ${C_p}$ is 1006.3 J/(kg K), and thermal conductivity $k$ is 0.02816 W/(m k), the initial temperature distribution ${T_0}$ is 300 K, and the gravitational acceleration $g$ is 9.81 m/s$^{ - 2}$. In this case, the dimensionless Rayleigh number $Ra$ ($ = {\rho _0}^2{C_p}g\beta \Delta T{({R_{outer}} - {R_{inner}})^3}/(k\mu )$), which represents the thermal behavior characteristics of the fluid, is 4.9$ \times {10^4}$ and the Prandtl number $Pr$ is 0.7147.

\begin{figure}
\begin{center}
\includegraphics[width=1.0\textwidth]{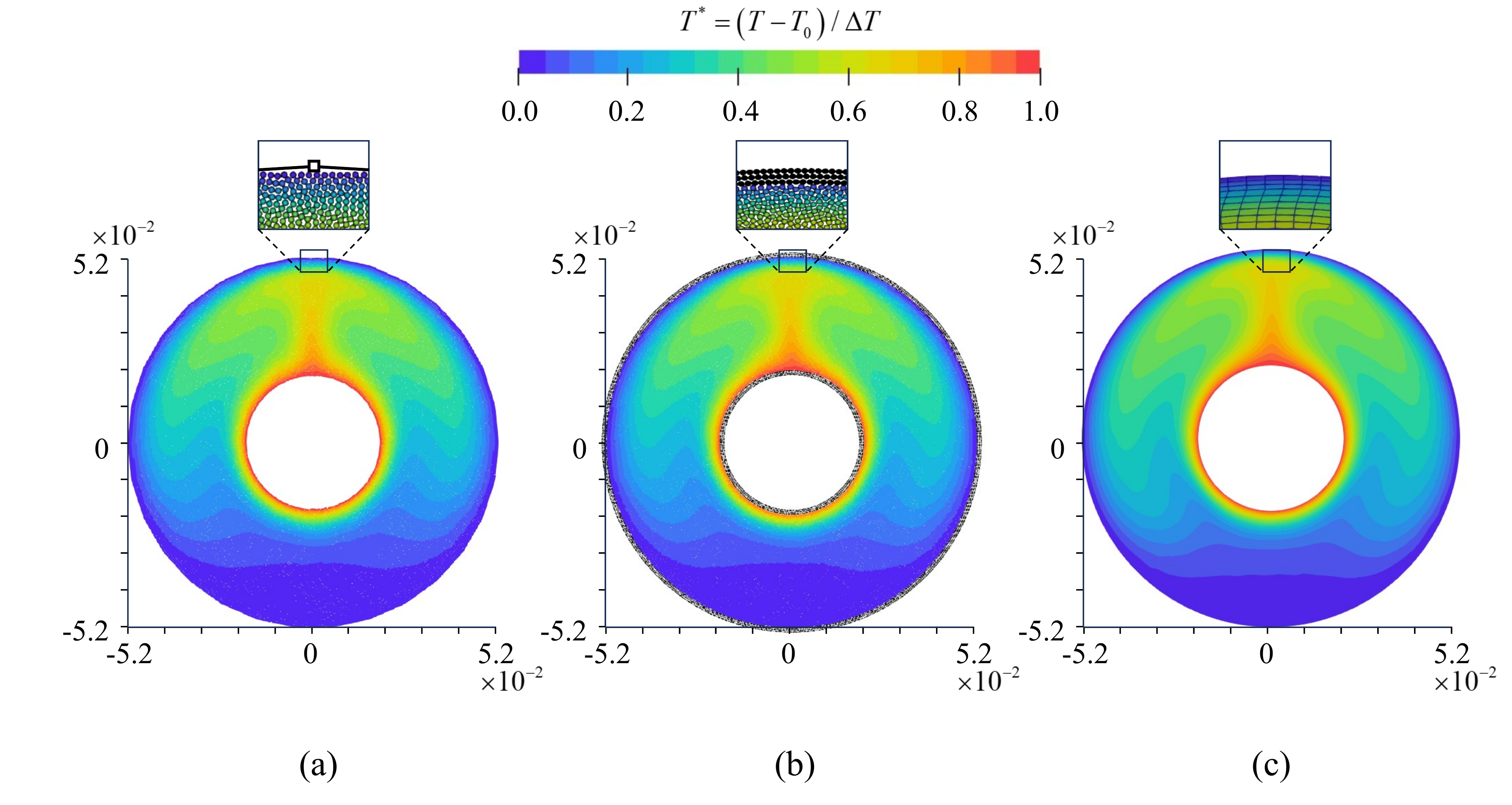}
\end{center}
\caption{Comparison of the temperature contours for natural convection in a horizontal concentric annulus problem among three methods: (a) SPH with segment-based boundary treatment method (proposed), (b) SPH with boundary particle method, and (c) FVM.
}\label{fig:conCylinder_contour_temp}
\end{figure}

\begin{figure}
\begin{center}
\includegraphics[width=1.0\textwidth]{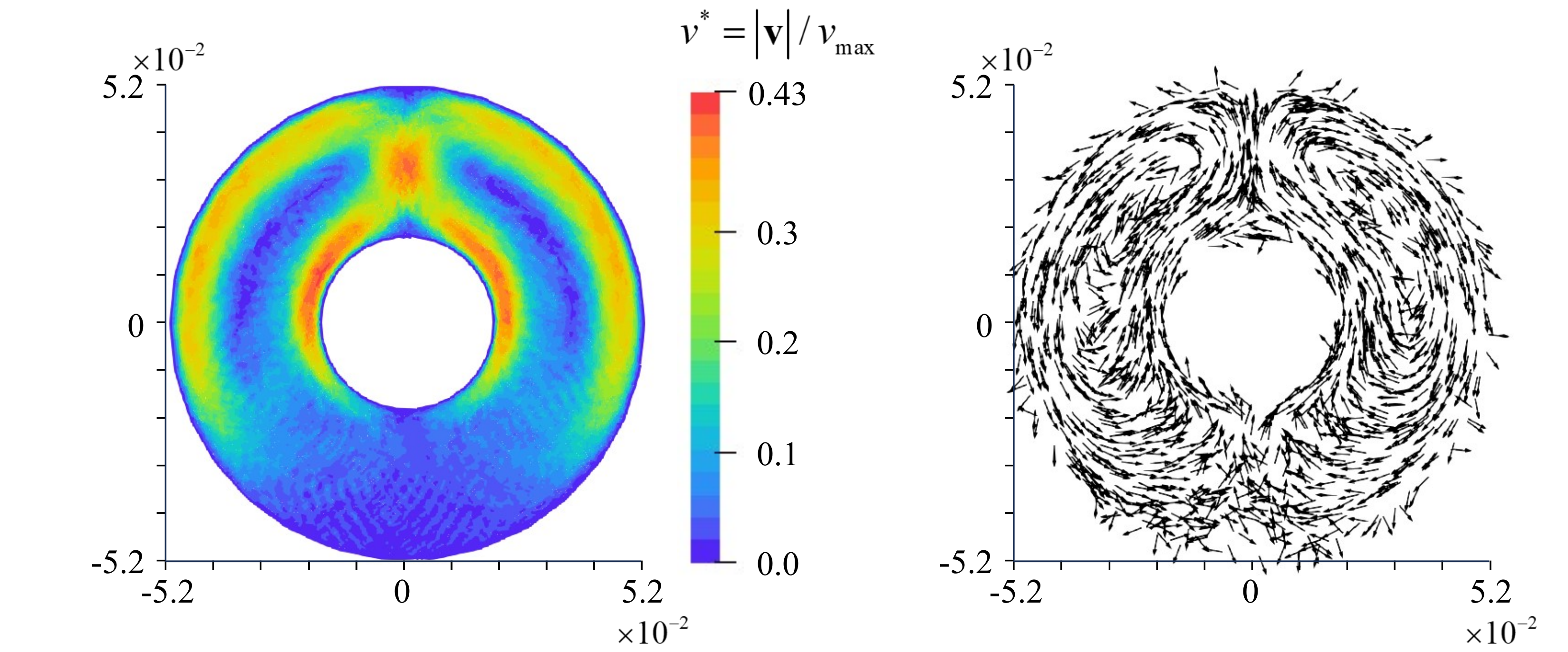}
\end{center}
\caption{Velocity field distribution from the results of the proposed method for natural convection in a horizontal concentric annulus problem: (a) velocity magnitude contour and (b) velocity vector distribution. Note that the velocity vector arrow size is not scaled to its magnitude.
}\label{fig:conCylinder_contour_vel}
\end{figure}

In the SPH simulation, the initial particle spacing $\Delta x$ is set to verify convergence according to particle resolution for the three cases of 0.002, 0.001, and 0.0005 m. The time step size $\Delta t$ for each case is set to 1$ \times {10^{ - 3}}$, 5$ \times {10^{ - 4}}$, and 2.5$ \times {10^{ - 4}}$ s, respectively. ${v_{\max }}$ is calculated as ${v_{\max }} = \sqrt {g\beta \Delta T({R_{outer}} - {R_{inner}})} $ and uses a value of 0.15. Figs. \ref{fig:conCylinder_seg_BPM}(a) and (b) show the particle discretization domain for the proposed method and the method using boundary particles, respectively. Here, $\Delta x$ is 0.0005 m. For the slip condition, a no-slip boundary condition is imposed on all wall boundaries.

In the FVM simulation, 26580 grids are used to model the problem domain, and the time step size is 2.5x10-4 s. It is simulated using an incompressible transient solver, the turbulence model is k-omega, and the state equation is the Boussinesq approximation. The detailed setting information is provided in Table \ref{tab:OpenFOAM_settings}.

\begin{figure}
\begin{center}
\includegraphics[width=1.0\textwidth]{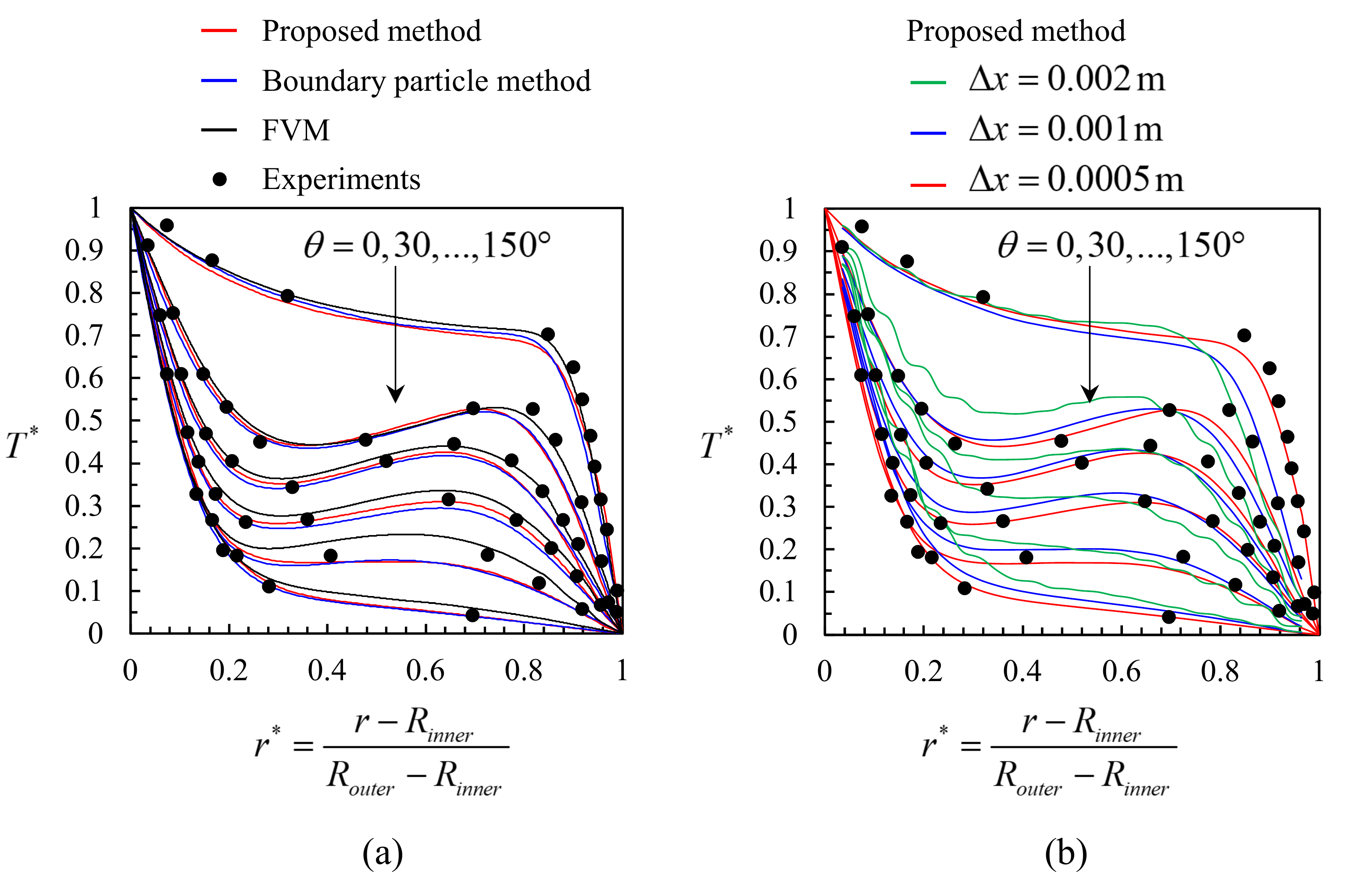} 
\end{center}
\caption{Non-dimensionalized temperature profile on different $\theta$ lines: (a) comparison between the proposed method, boundary particle method, FVM and experiments, and (b) convergence test at $dx$=0.002, 0.001, and 0.0005 m.
}\label{fig:conCylinder_results}
\end{figure}

\fref{fig:conCylinder_contour_temp}(a) illustrates the temperature contour of the proposed method for the case where $\Delta x$ is 0.0005 m at $t$= 20 s. The heat of the fluid adjacent to the inner hot wall gradually transfers outward, resulting in buoyancy owing to the temperature difference. As shown in \fref{fig:conCylinder_contour_vel}, the fluid particles heated by the inner cylinder rise, cool upon reaching the outer cylinder, and then descend along the outer perimeter, creating an overall circulation. As shown in Figs. \ref{fig:conCylinder_contour_temp}(a)-(c), the results of the proposed method exhibit a favorable alignment with the SPH results using boundary particles and FVM simulations.

For this benchmark problem, the results can be quantitatively verified using the experimental data from \cite{kuehnExperimentalStudyNatural1978}. In the problem domain description in Fig. 11, the $\theta $ value is increased by 30 \textdegree, and the non-dimensionalized temperature T* is measured as $(T - {T_{cold}})/\Delta T$. \fref{fig:conCylinder_results}(a) compares the proposed method with the boundary particle method, FVM, and experimental results, revealing good agreement. \fref{fig:conCylinder_results}(b) shows the results of the convergence test. The proposed method shows a tendency to agree well with the results obtained using the boundary particle method, FVM, and experiments. In the convergence test, the oscillation of the temperature profile is observed when $\Delta x$ = 0.002 m. In SPH, the temperature at the probe point is obtained by referring to the temperature values of surrounding particles \cite{englishModifiedDynamicBoundary2022}. A coarse particle resolution can lead to large temperature differences among the referenced particles at the probe point, resulting in oscillations.

\subsection{Complex shaped cavity}

\begin{figure}
\begin{center}
\includegraphics[width=0.95\textwidth]{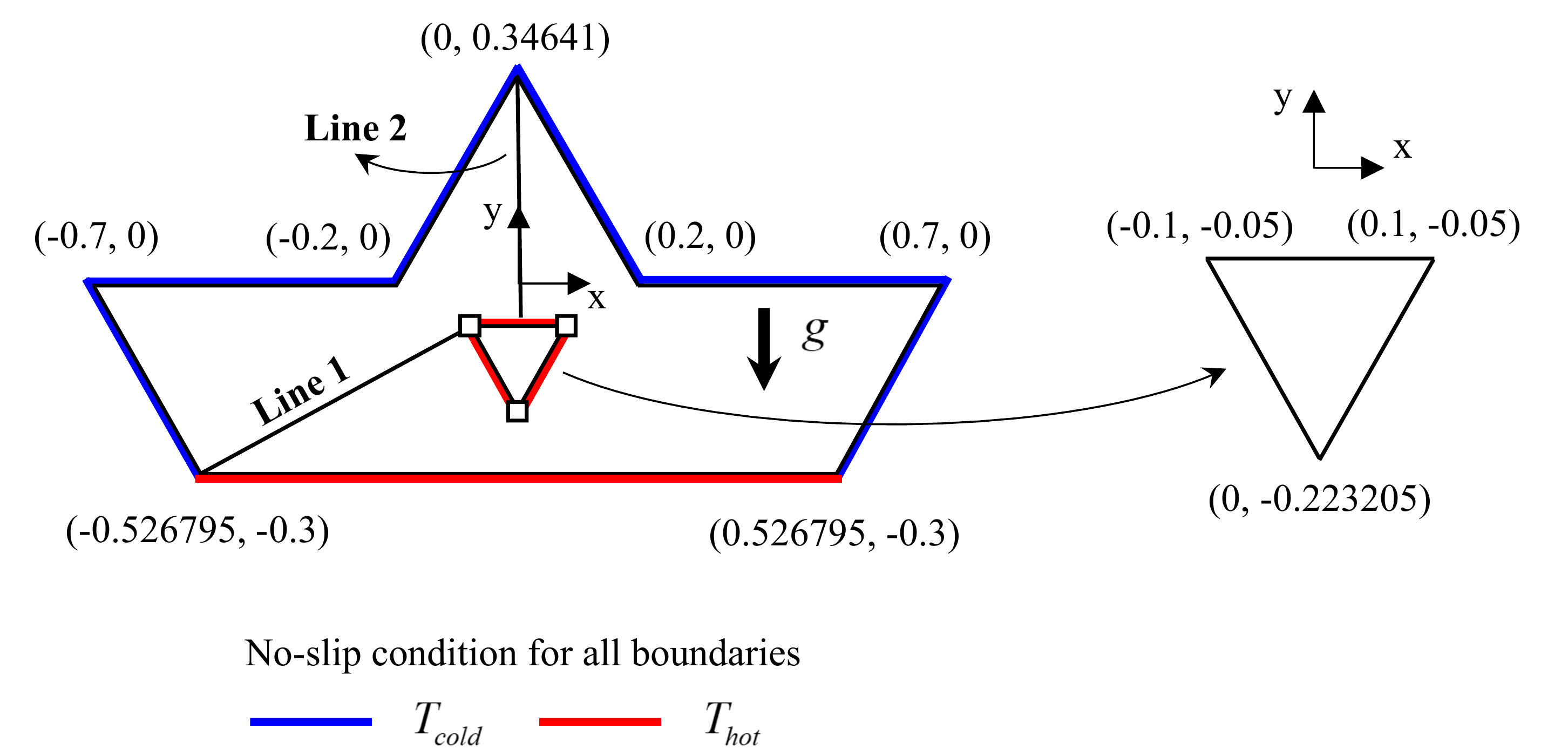} 
\end{center}
\caption{Domain description for a complex shaped cavity problem.
}\label{fig:complexCavity_domain}
\end{figure}

\begin{figure}
\begin{center}
\includegraphics[width=0.95\textwidth]{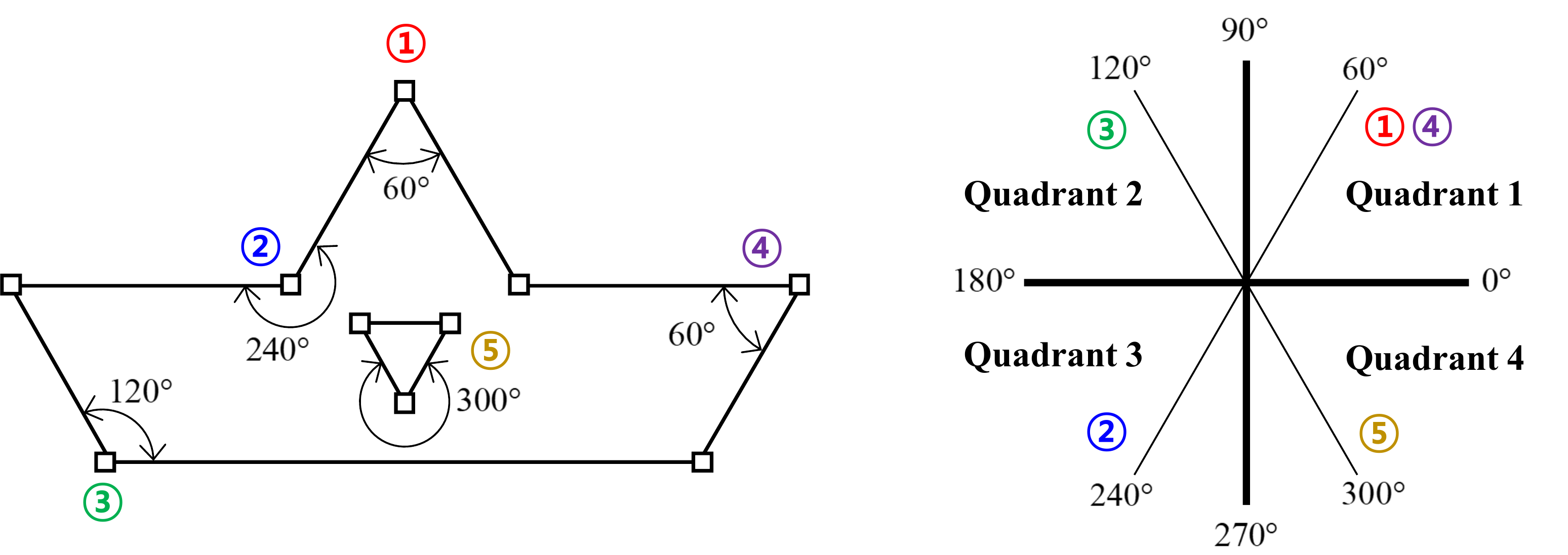} 
\end{center}
\caption{Angle distribution between boundary segments for a complex shaped cavity problem.
}\label{fig:complexCavity_angle_graph}
\end{figure}

\begin{figure}
\begin{center}
\includegraphics[width=0.53\textwidth]{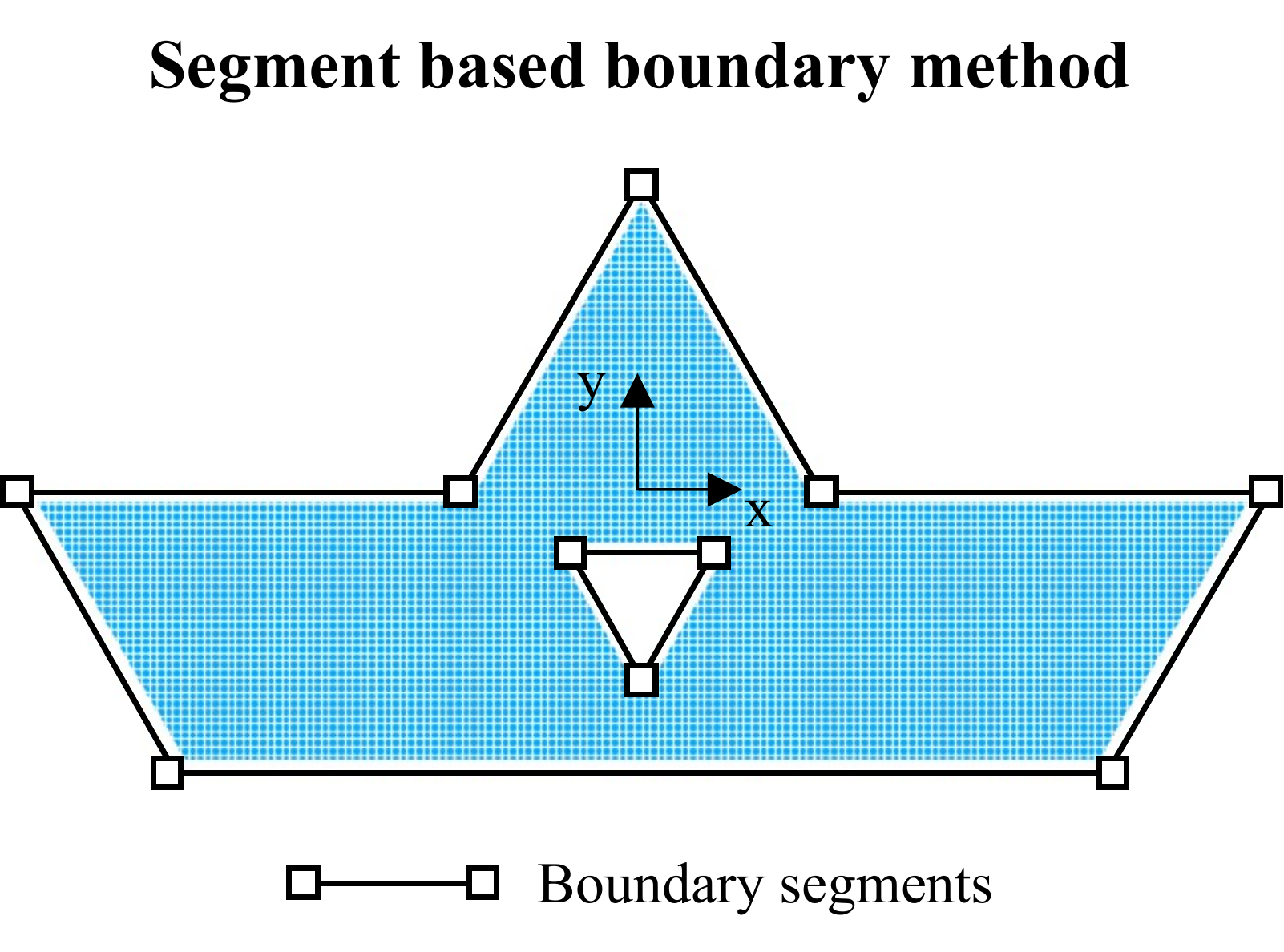}
\end{center}
\caption{Particle discretization domain for a complex shaped cavity problem using segment-based boundary treatment method.
}\label{fig:complexCavity_segBoundary}
\end{figure}

\begin{figure}
\begin{center}
\includegraphics[width=1\textwidth]{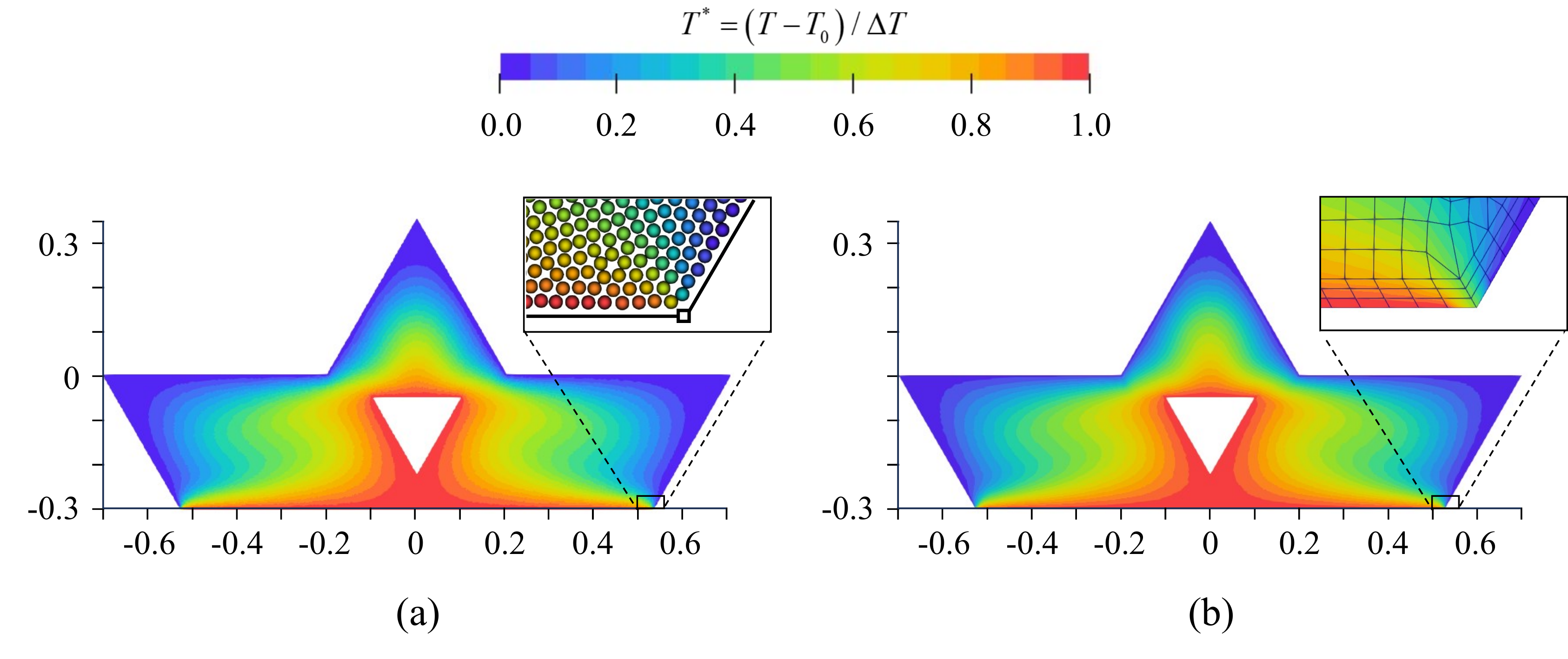} 
\end{center}
\caption{Comparison of the temperature contours for complex shaped cavity problem between two methods: (a) SPH with segment-based boundary treatment method (proposed) and (b) FVM.
}\label{fig:complexCavity_contour_temp}
\end{figure}

\begin{figure}
\begin{center}
\includegraphics[width=1\textwidth]{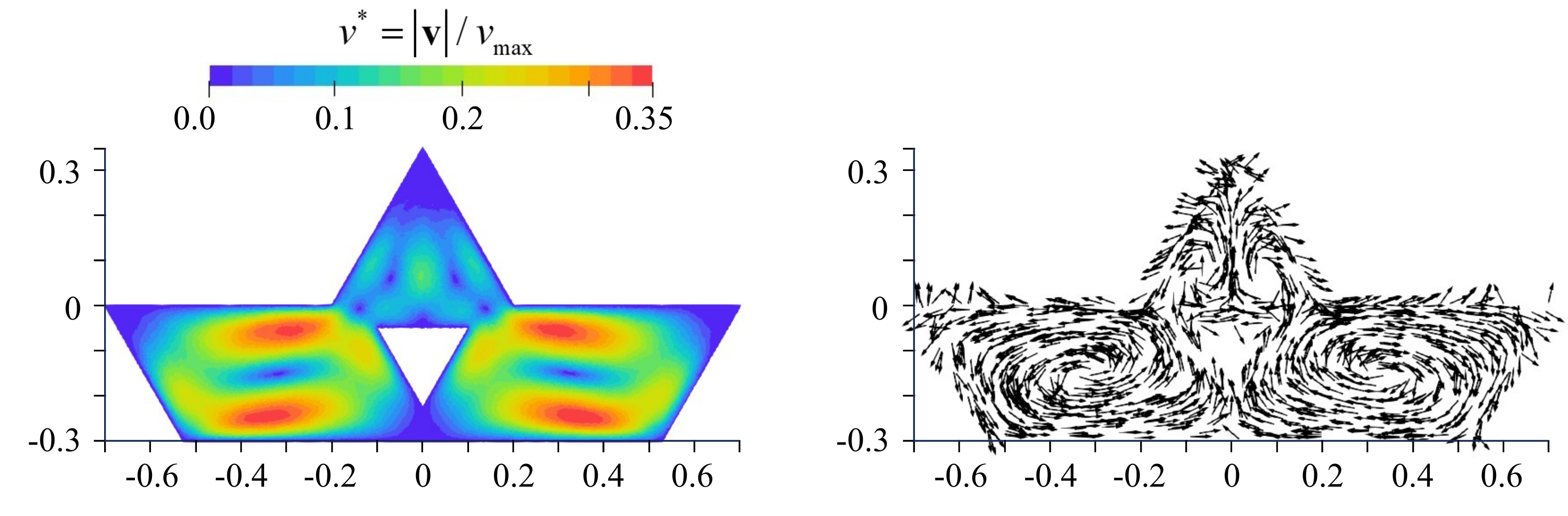} 
\end{center}
\caption{Velocity field distribution from the results of the proposed method for a complex shaped cavity problem: (a) velocity magnitude contour and (b) velocity vector distribution. Note that the velocity vector arrow size is not scaled to its magnitude.
}\label{fig:complexCavity_contour_vel}
\end{figure}

\begin{figure}
\begin{center}
\includegraphics[width=0.95\textwidth]{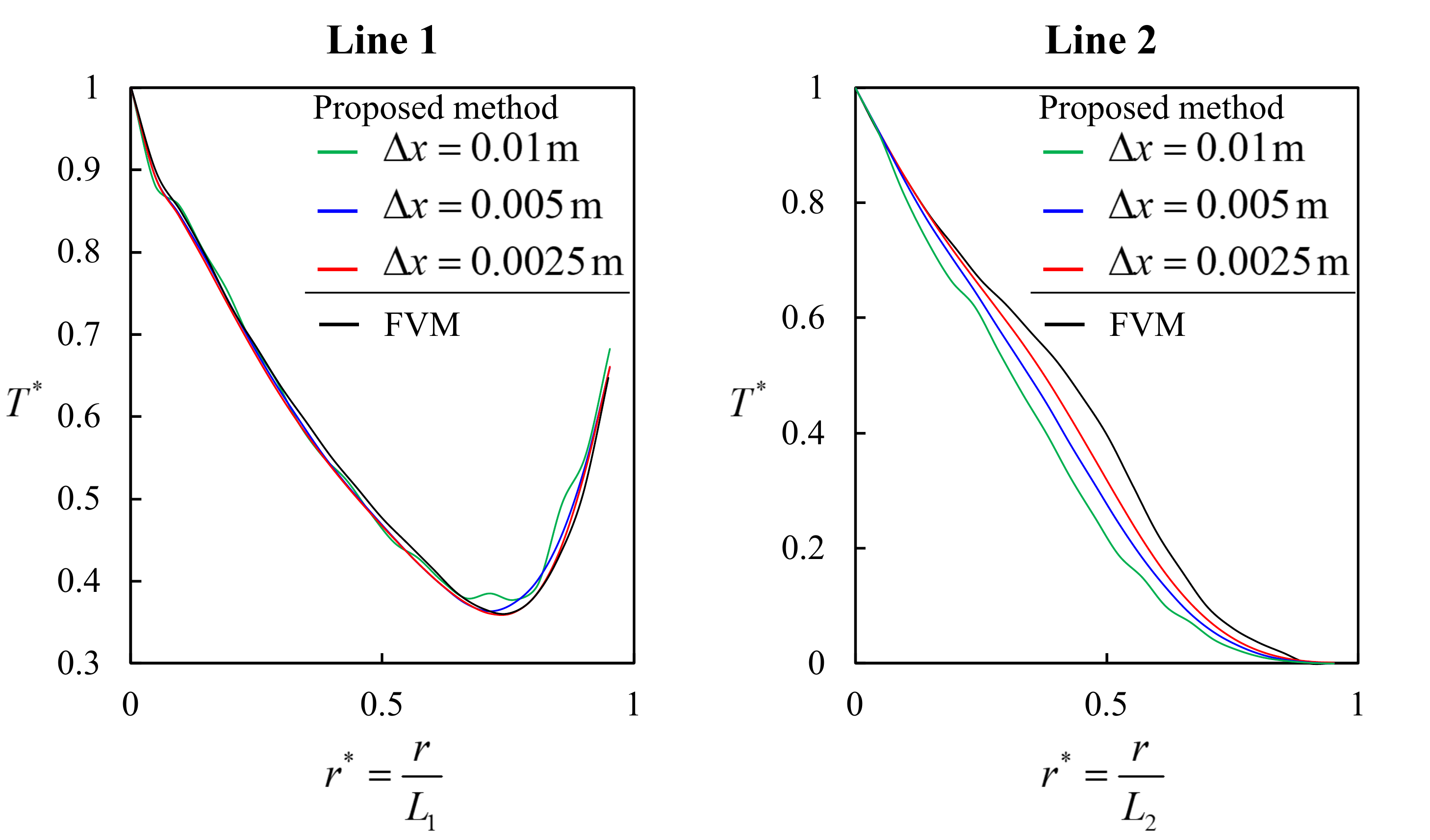} 
\end{center}
\caption{Comparison of the non-dimensionalized temperature profiles between the proposed method and FVM, and convergence tests at $dx$=0.01, 0.005, and 0.0025 m: (a) results along line1 and (b) results along line2.
}\label{fig:complexCavity_results}
\end{figure}

As a final example, a flow analysis is performed in a cavity with a complex shape, as shown in \fref{fig:complexCavity_domain}. The proposed method offers greater modeling flexibility than the boundary particle method. While modelling complex shapes is relatively easy, demonstrating that the solution remains accurate and stable for such complexities is eqaully important. To account for the influence of the geometrc complexity, the flow distribution must exhibit stability at both concave and convex corners. As shown in \fref{fig:complexCavity_angle_graph}, the shape is designed such that the angles formed between the line segments are distributed across all four quadrants of the angle graph. The simulation results are verified by comparison with the FVM results.

The domain consists of outer boundary and inner boundaries. The outer boundary consists of the following seven nodes: (0, 0.34641), (-0.2, 0), (-0.7, 0), (-0.526795, -0.3), (0.526795, -0.3), (0.7, 0), and (0.2, 0). The internal boundary consists of the following 3 nodes: (-0.1, -0.05), (0.1, -0.05), and (0, -0.223205), see \fref{fig:complexCavity_domain}. The unit for all node positions is meters. The temperature of the outer boundary, excluding the bottom wall, is 300.0 K, while that of the bottom wall is 300.005 K. The temperature of the inner boundary is set to 300.005 K, so that the maximum temperature difference between the boundaries is $\Delta T$= 0.005 K. The initial density of the internal fluid ${\rho _0}$ is 1.0 kg/ m$^3$, dynamic viscosity $\mu $ is 2.0$ \times {10^{ - 5}}$ Pa s, thermal expansion coefficient $\beta $ is 0.003 K$^{ - 1}$, specific heat capacity ${C_p}$ is 1000 J/(kg K), thermal conductivity $k$ is 0.028 W/(m k), the initial temperature distribution ${T_0}$ is 300 K, and the gravitational acceleration $g$ is 9.81 m/s$^{ - 2}$. The reference length ${L_{ref}}$ to define ${v_{\max }}$ is set to 0.3 m. In this case $Ra$ is 7.1$ \times {10^3}$ and $Pr$ is 0.7143.

In the SPH simulation, three cases, with initial particle spacings $\Delta x$ of 0.01, 0.005, and 0.0025 m, are considered, and the time step sizes $\Delta t$ for each case are set to 0.1, 0.05, and 0.02 s, respectively. ${v_{\max }}$ is calculated as ${v_{\max }} = \sqrt {g\beta \Delta T{L_{ref}}} $ and is 0.0066 m/s. \fref{fig:complexCavity_segBoundary} shows the particle discretization domain for the proposed method when $\Delta x$= 0.0025 m. For the slip condition, a no-slip boundary condition is imposed on all wall boundaries.

In the FVM simulation, 18994 grids are used to model the problem domain, and the time step size is 0.02 s. It is simulated using an incompressible transient solver, the turbulence model is k-omega, and the state equation is the Boussinesq approximation. The detailed setting information is provided in Table \ref{tab:OpenFOAM_settings}.

\fref{fig:complexCavity_contour_temp}(a) shows the temperature contour of the proposed method for the $\Delta x$= 0.0025 m case at $t$= 1000 s. Buoyancy occurs depending on the temperature difference between each wall, which causes the circulation of internal flow (\fref{fig:complexCavity_contour_vel}). The results of the proposed method match well with the FVM results, as shown in Figs. \ref{fig:complexCavity_contour_temp}(a) and (b).

To examine the accuracy of the proposed method, a comparison with FVM is performed on the temperature distribution along the two lines with the largest thermal change rate in the domain. Line 1 is defined by the following two positions: (-0.1, -0.05) and (-0.526795, -0.3). Line 2 is defined by the following two positions: (0, -0.05) and (0, 0.34641), see \fref{fig:complexCavity_domain}. Here, the unit of position is meters. In \fref{fig:complexCavity_results}, the results of the two methods (the proposed method and FVM) are in good agreement, and the proposed method accurately simulates heat transfer from the wall boundary to the internal fluid. Note that the $r$ is the distance from each line's initial position to end position, and $L_{1}$ and $L_{2}$ denote the lengths of line 1 and line 2, respectively.

\section{Conclusion}
\label{sec:conclusion}

In this study, we proposed an SPH model to simulate the natural convection induced by heat transfer from wall boundaries. Traditionally, SPH mainly uses multi-layer boundary particles for wall boundary modeling, but this has limitations such as complexities in boundary modeling and increased computational cost. To solve this issues, the segment-based boundary treatment method has recently attracted attention. This study aimed to develop a wall heat transfer model based on this method.

In SPH, wall boundary modeling through segments was performed using lines in 2D and faces in 3D, offering the advantage of easy applicability, even for complex boundary shapes. Unlike the boundary particle method, which requires the entire boundary modeling with a resolution equivalent to the fluid particle size even for simple shapes, the segment-based approach only requires the modeling of segments where the normal vector of the boundary shape differs, leading to reduced computational costs. The model was derived through a mathematically rigorous derivation process, to maintain accuracy when using segments instead of boundary particles. The validity of the proposed method was confirmed through a comparison with the available experimental results, results obtained using boundary particles in SPH, and results from FVM.

SPH is widely used in the industry and academia because of its unique characteristics. However, there is substantial room for further development in terms of accuracy and practicality. The methodology presented in this paper holds the potential for practical application in a broader array of problems. A limitation of this study is the need to explore more diverse boundary conditions, such as the moving boundary and Neumann boundary conditions. In future work, we aim to overcome these limitations and extend the proposed method to three-dimensions for extensive applications.

\section*{Declaration of competing interest}
The authors declare that they have no known competing financial interests or personal relationships that could have appeared to influence the work reported in this paper.

\section*{Acknowledgement}
This research was also supported by Basic Science Research Program through the National Research Foundation of Korea (NRF) funded by the Ministry of Education. (2022R1C1C2006328).

\printbibliography

\end{document}